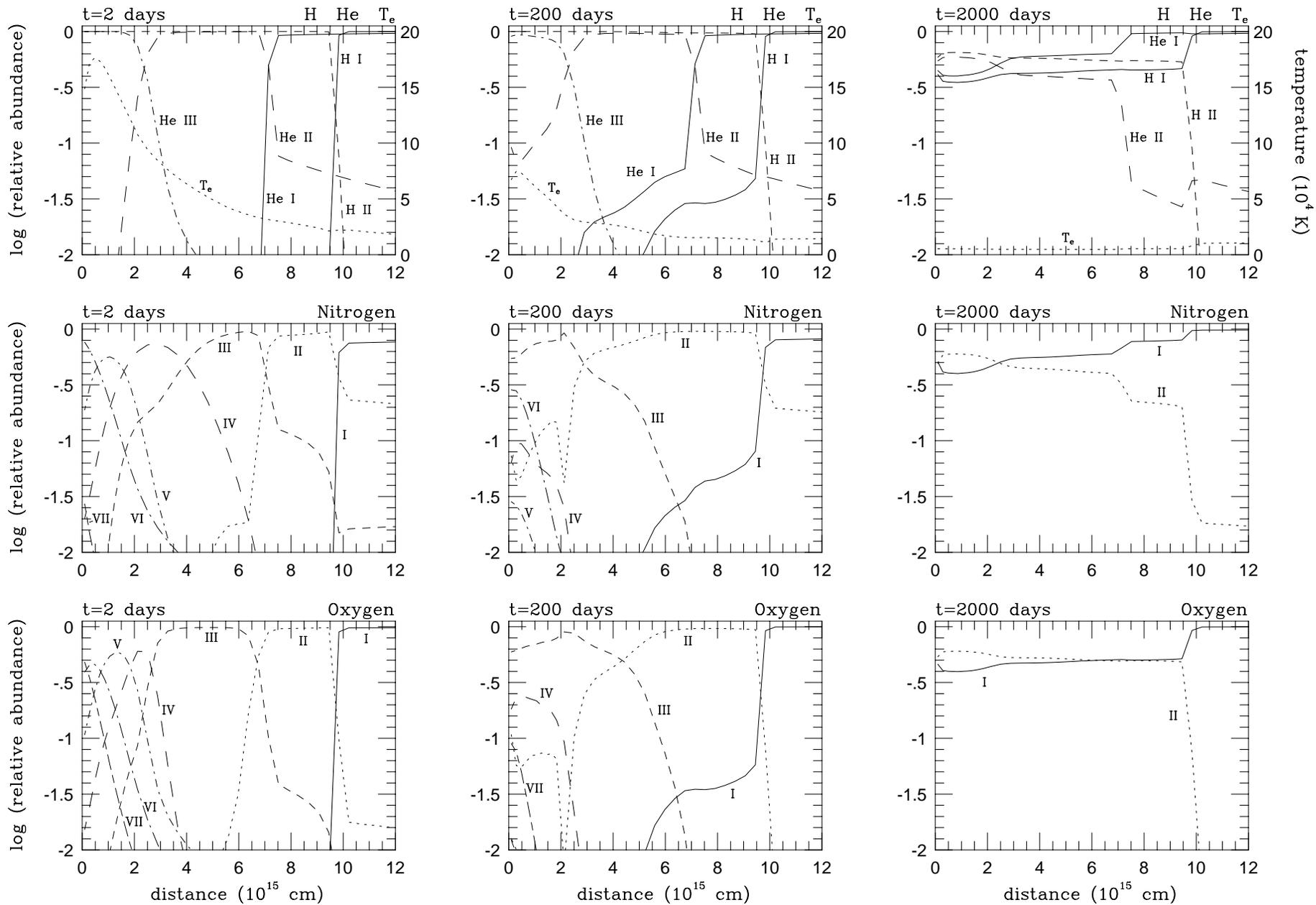

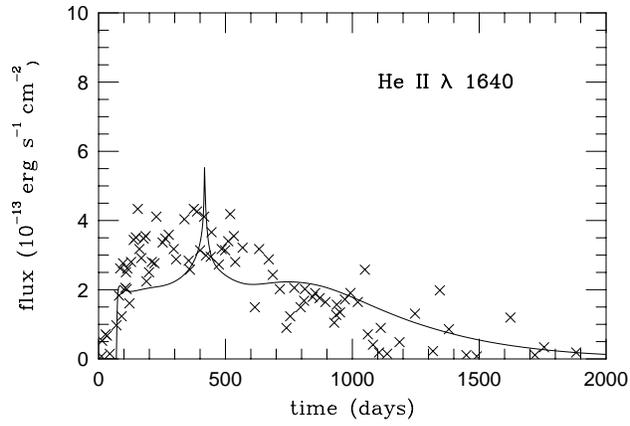
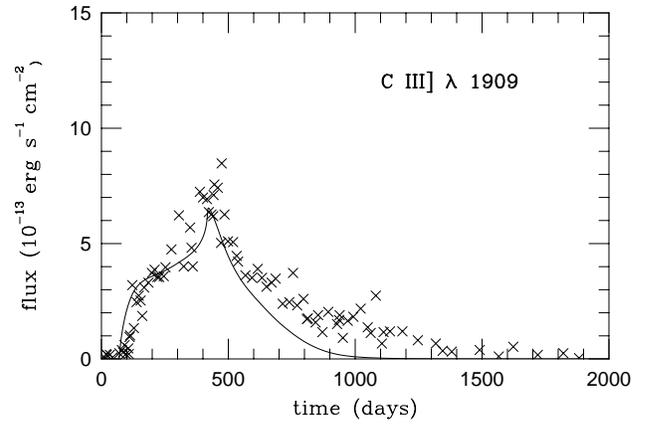
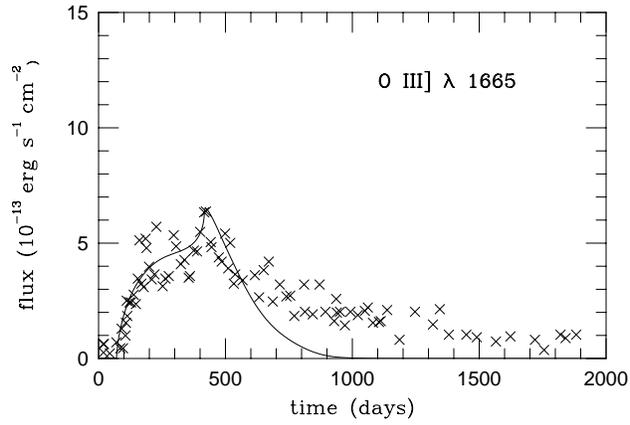
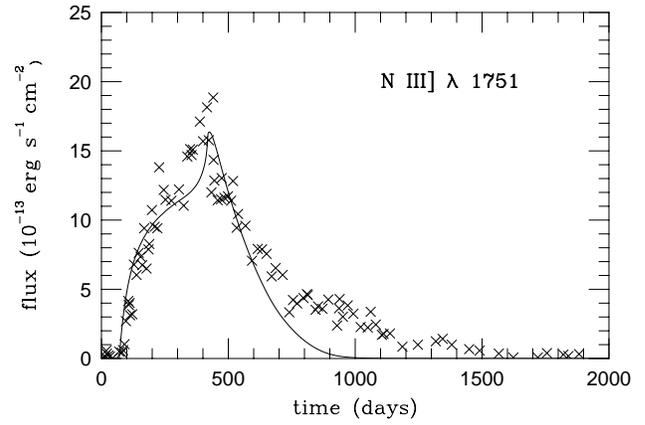
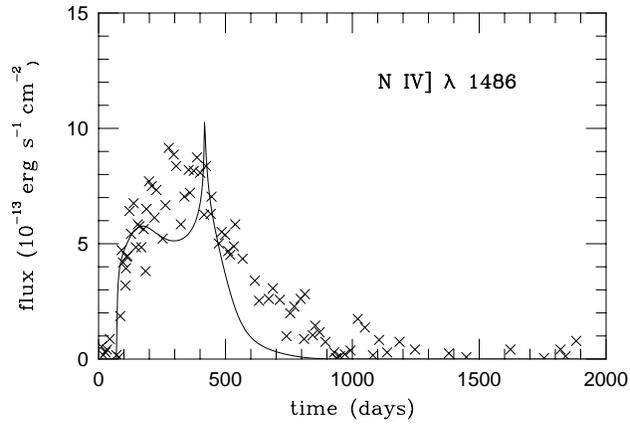
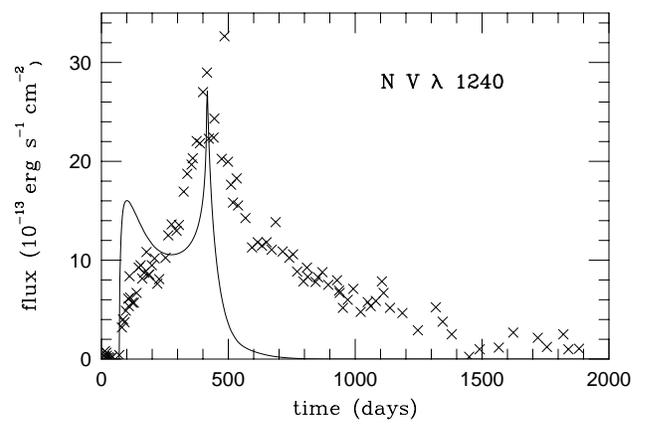
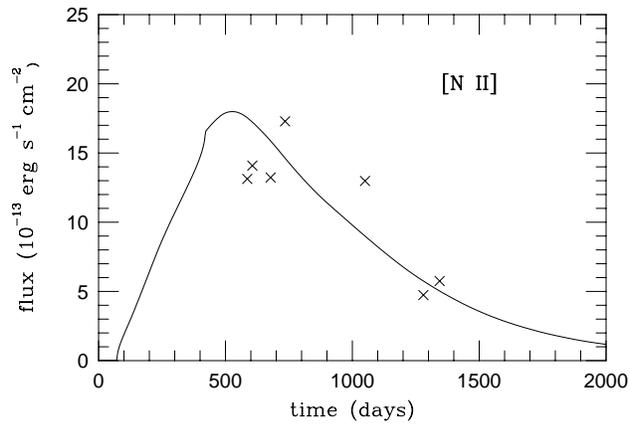
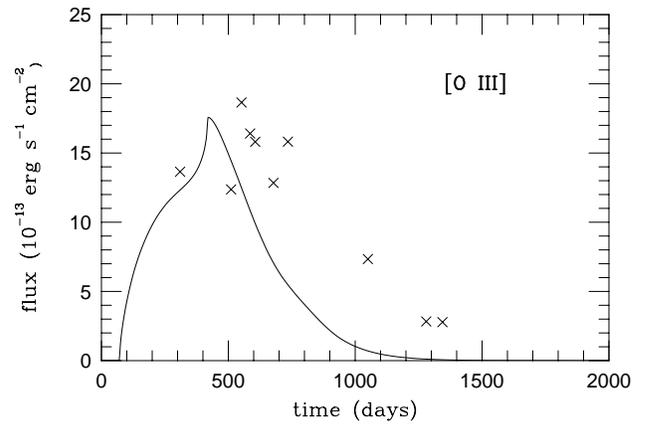

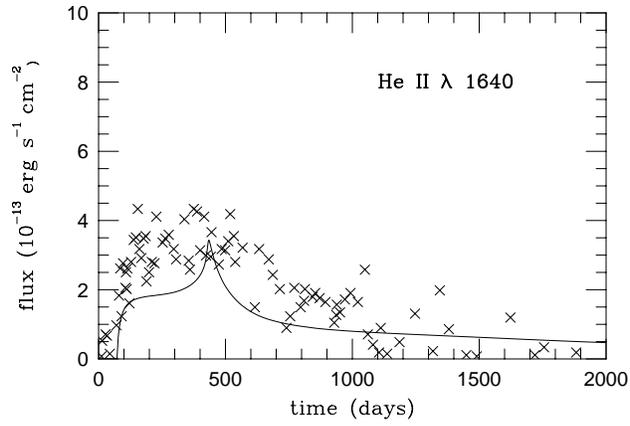
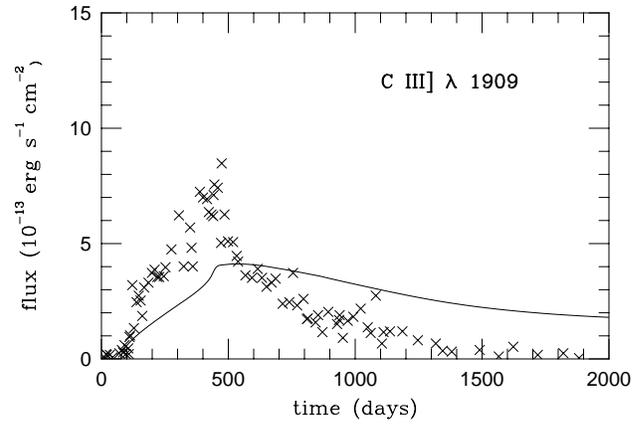
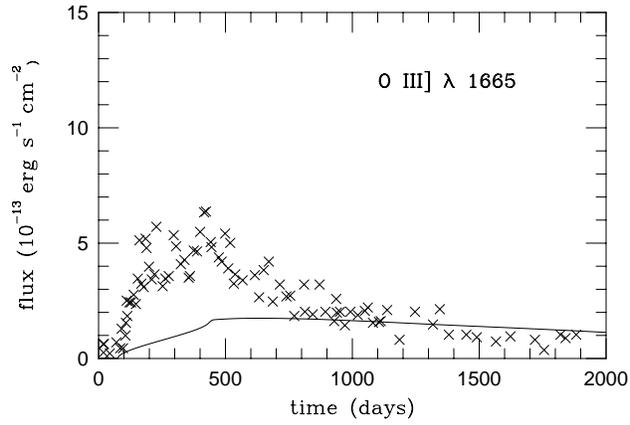
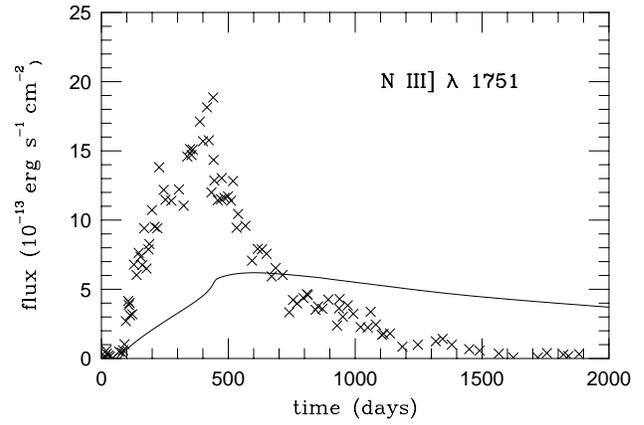
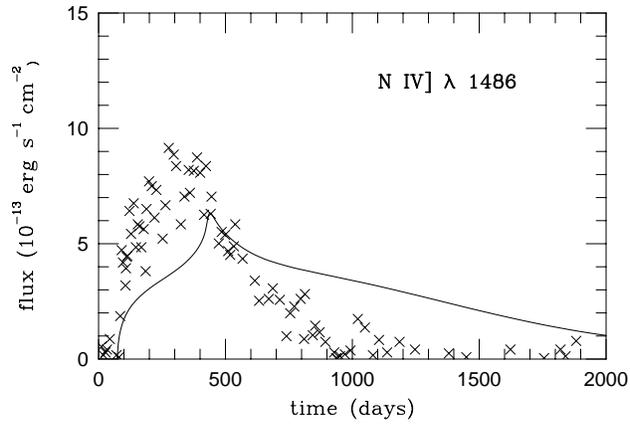
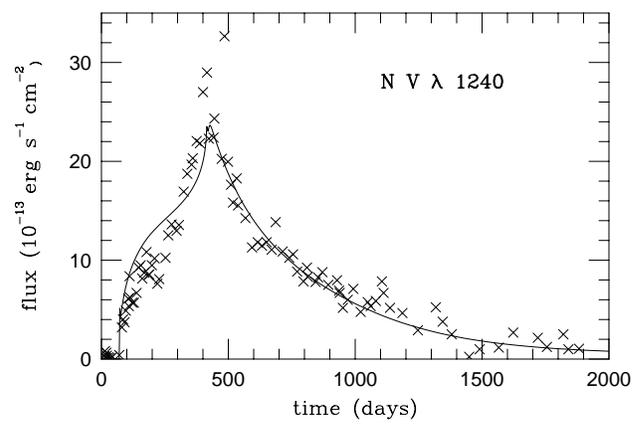
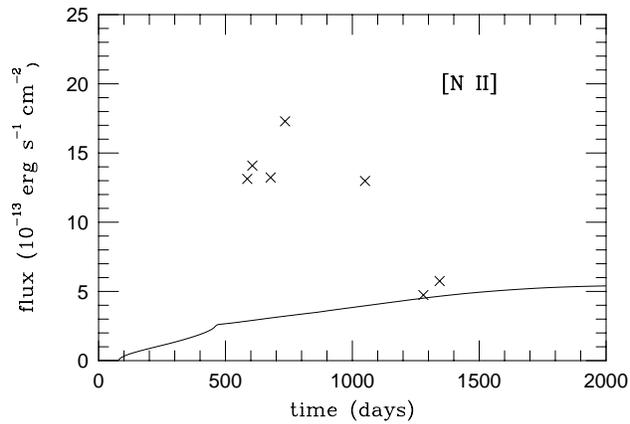
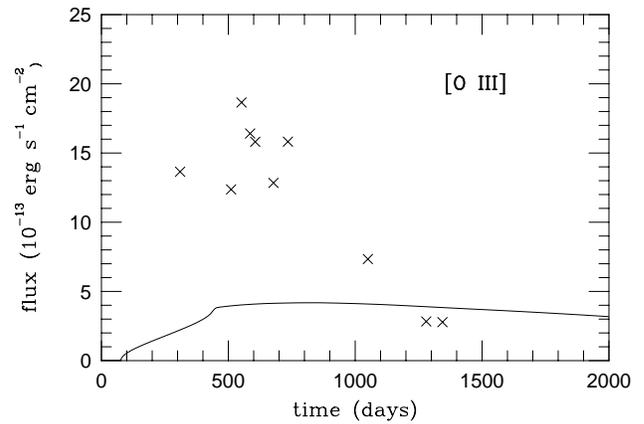

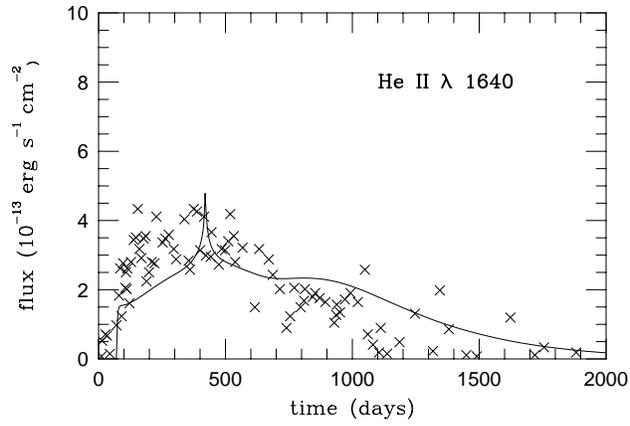
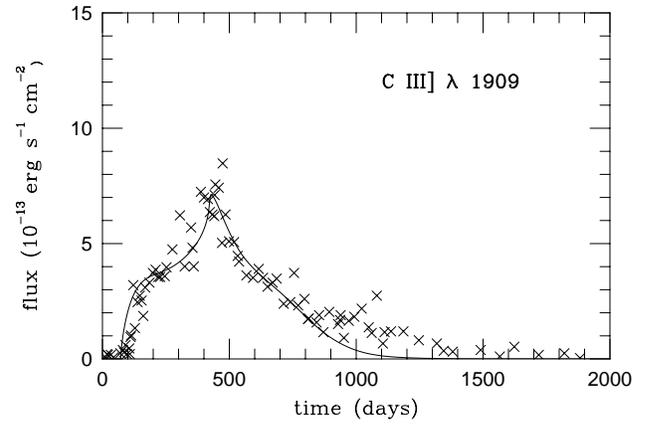
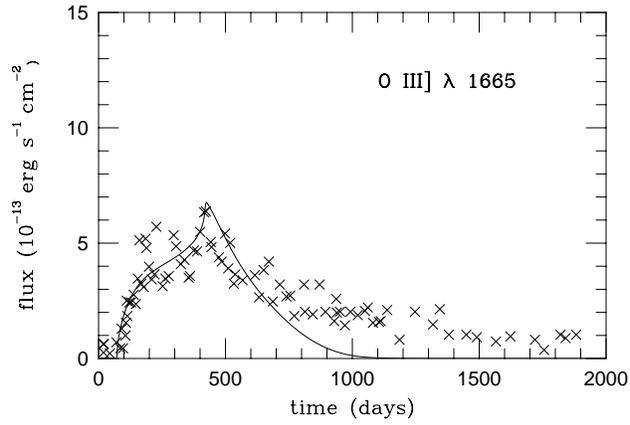
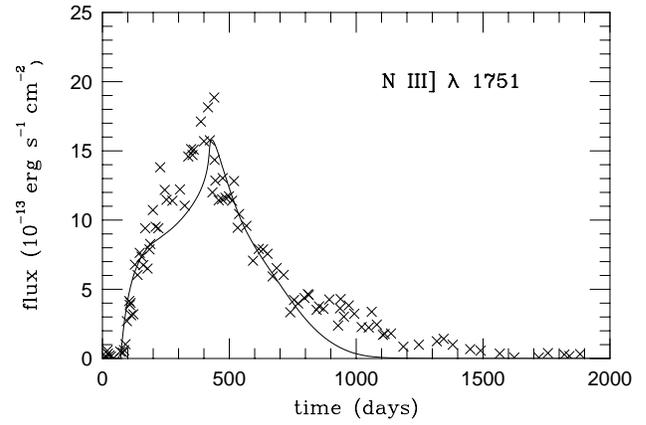
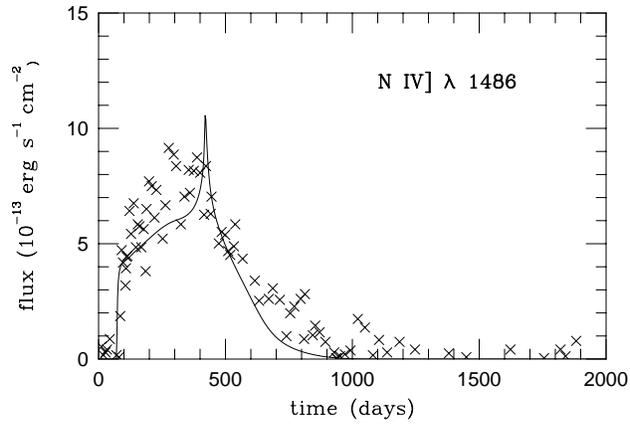
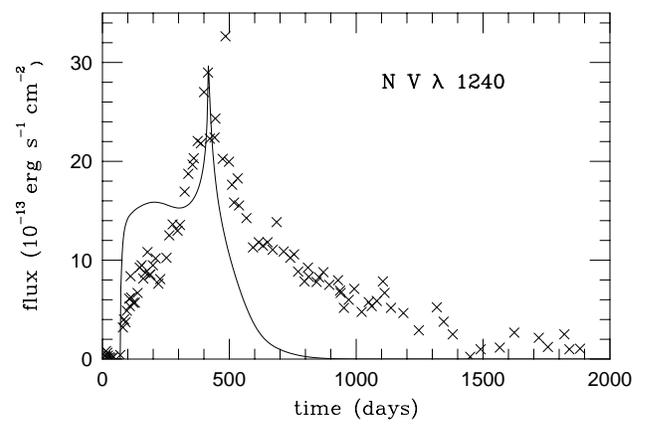
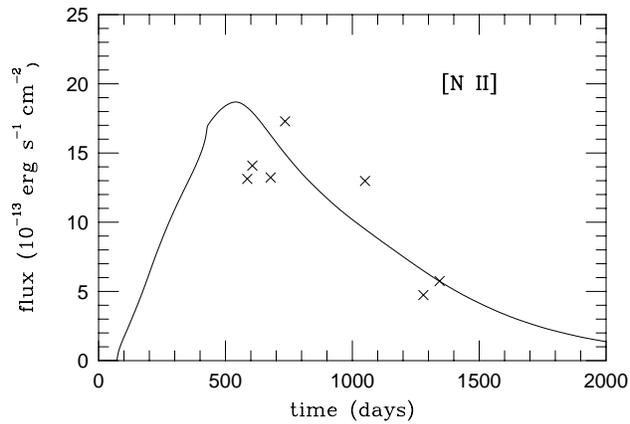
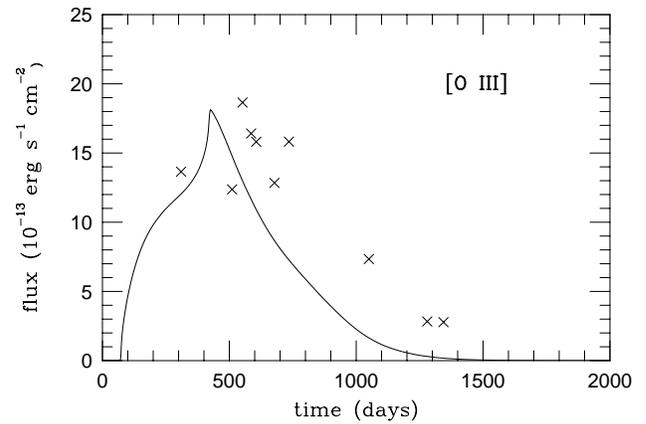

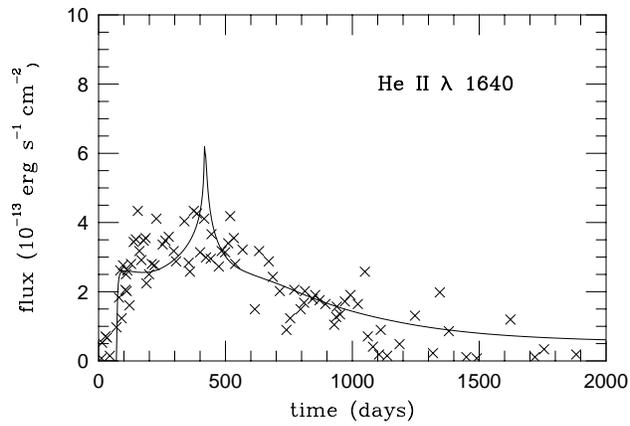
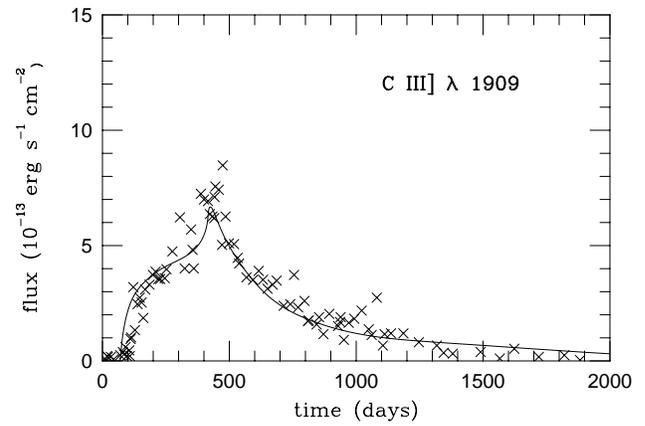
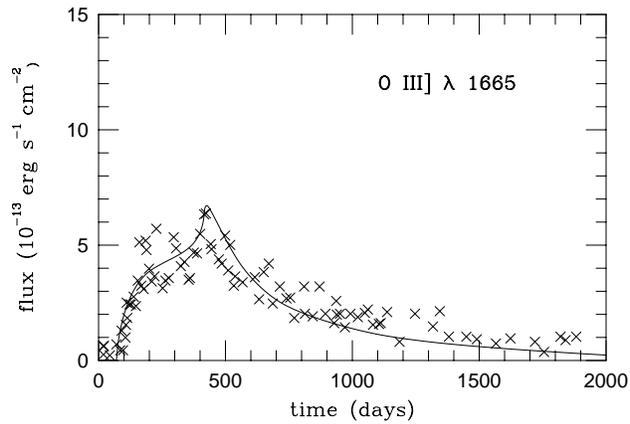
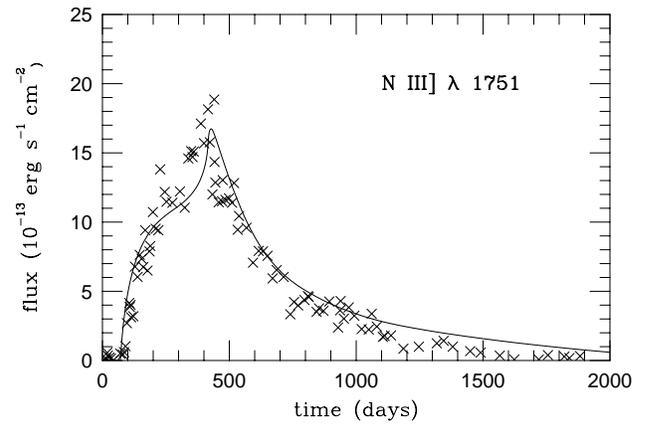
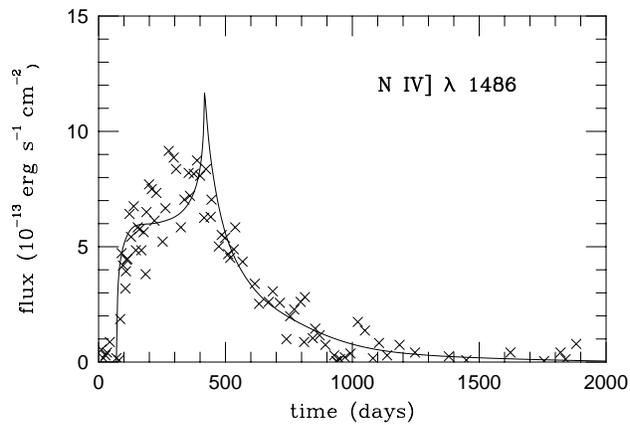
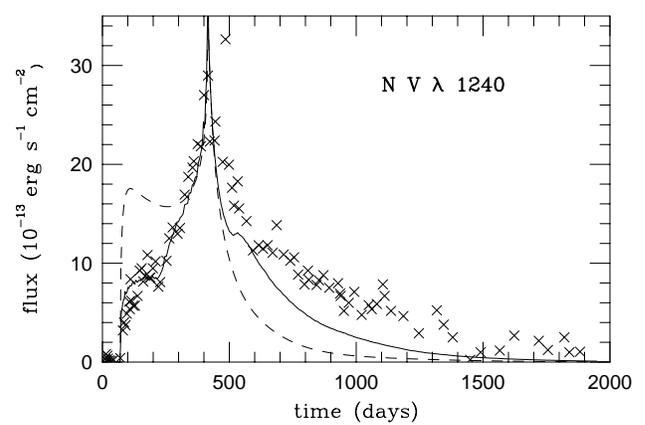
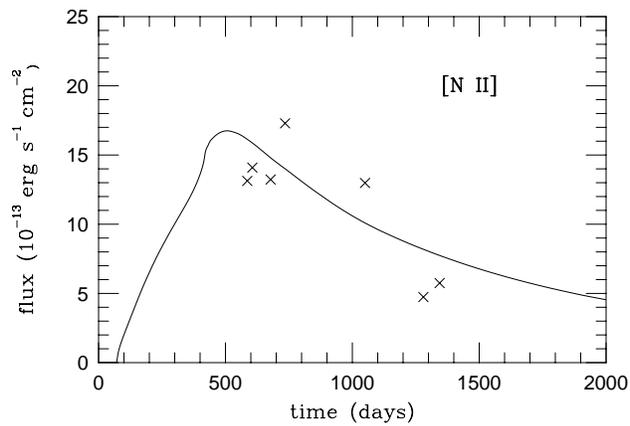
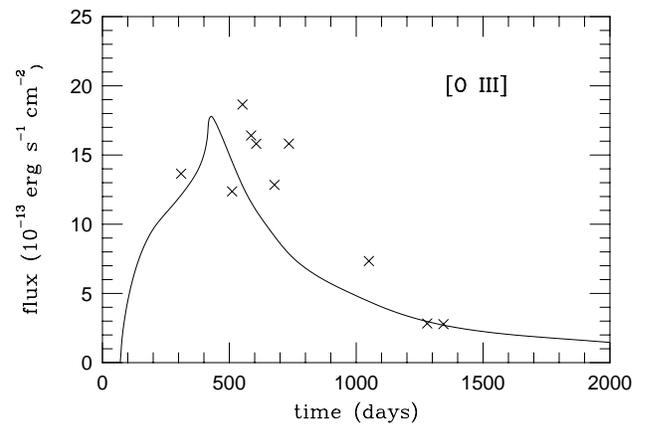

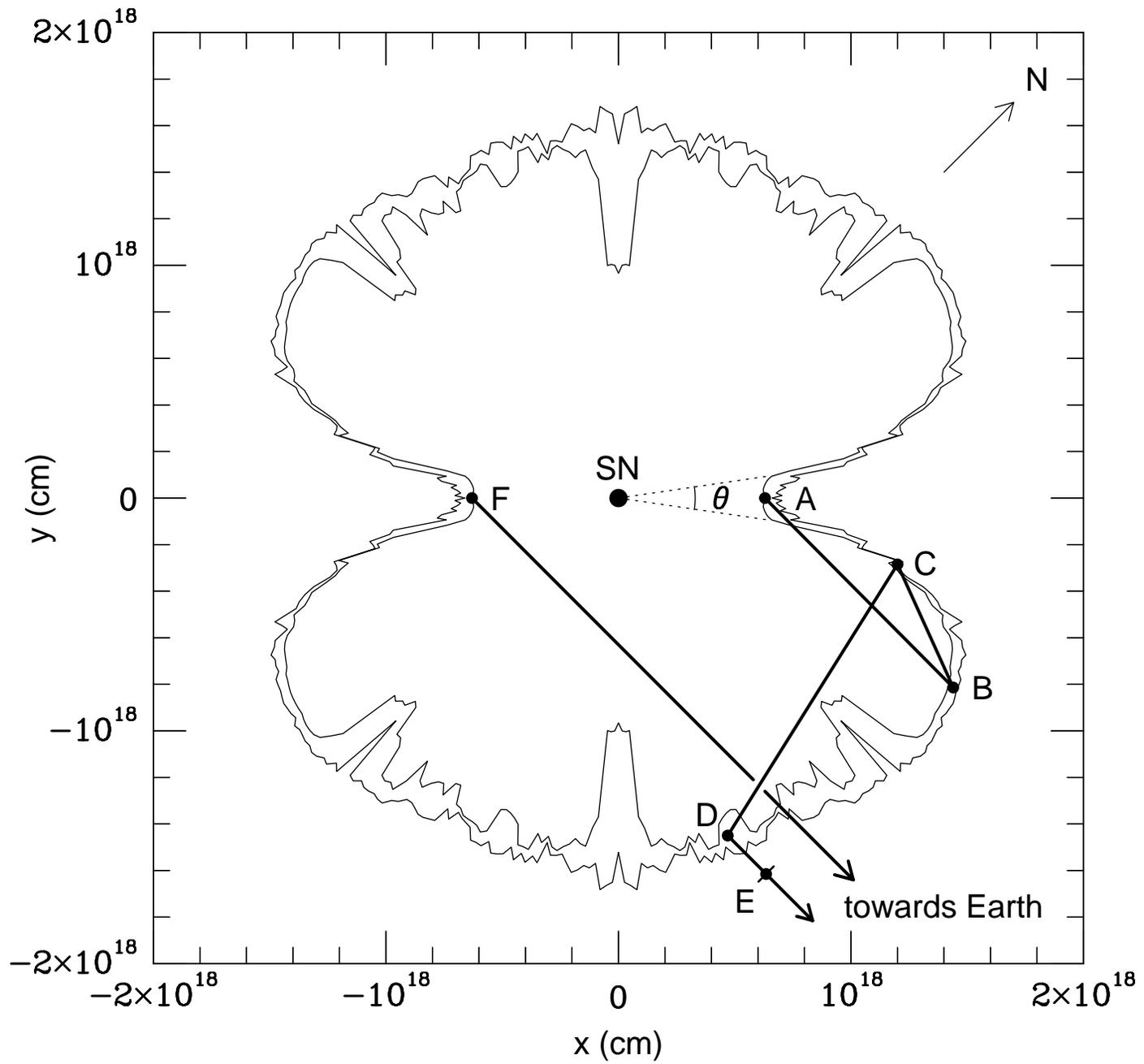

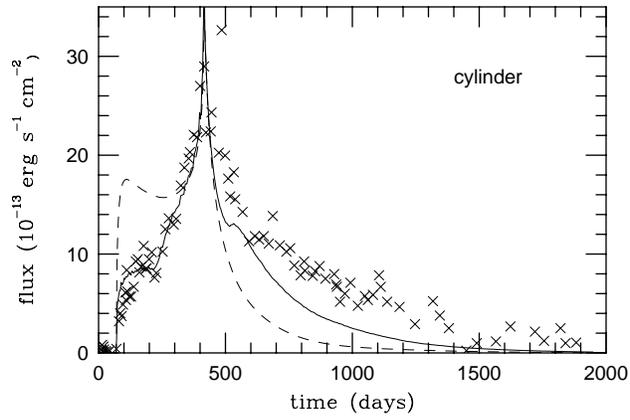
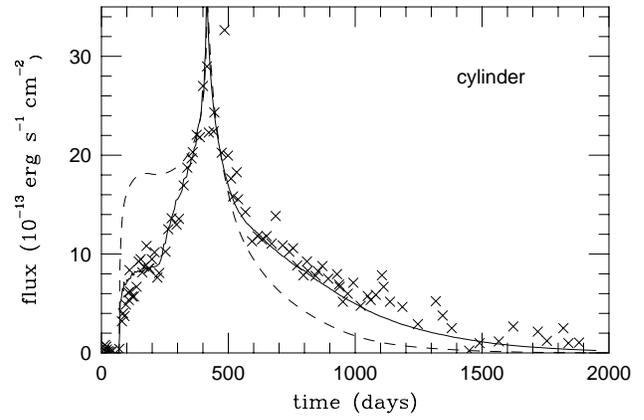
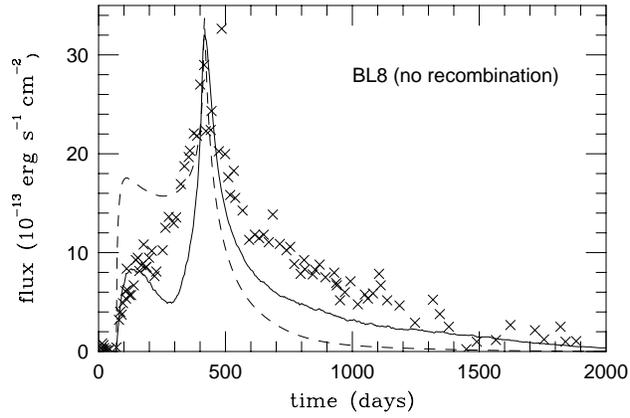
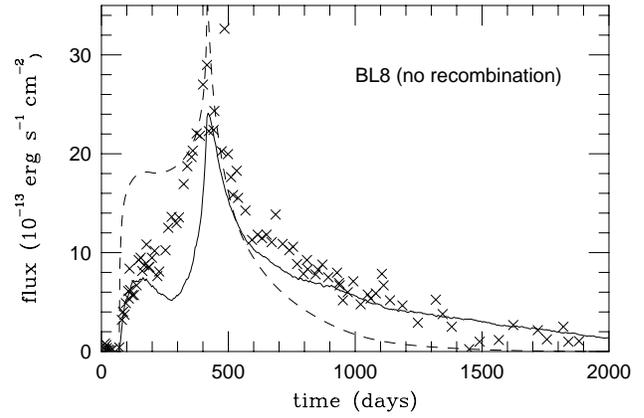
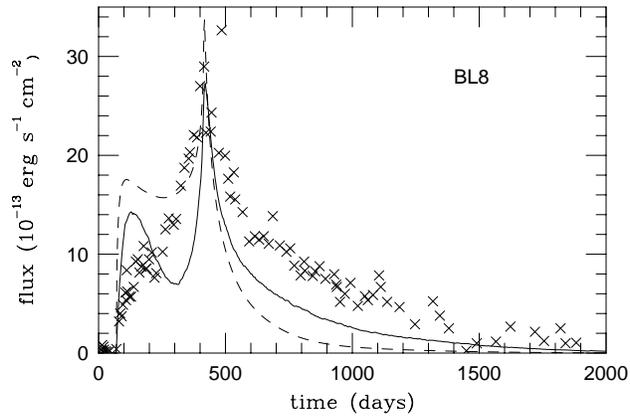
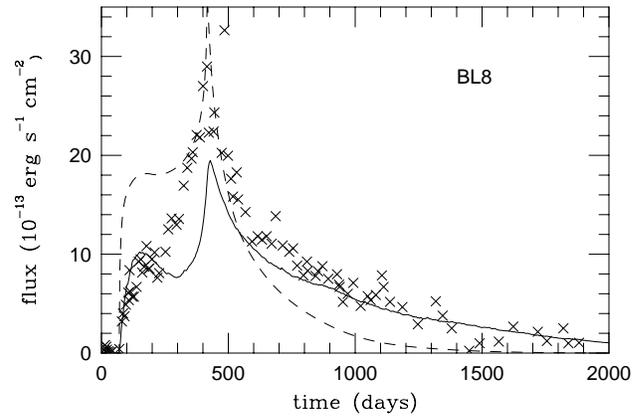

# THE LINE EMISSION FROM THE CIRCUMSTELLAR GAS AROUND SN 1987A

Peter Lundqvist[1] and Claes Fransson[2]
Stockholm Observatory, S-133 36 Saltsjöbaden, Sweden

## ABSTRACT

The narrow emission lines from the inner ring around SN 1987A during the first $\sim 2000$ days are modeled assuming that the ring was ionized by radiation accompanying the supernova breakout. The analysis extends that in Lundqvist & Fransson (1991) to include an improved description for the geometry, a multi-density structure of the emitting gas, results from improved calculations for the evolution of the EUV radiation from the outburst, and updated atomic data. To model the light curves of the lines, the ring has to be optically thick to the ionizing radiation, as is the case for a ring geometry, but not for a spherical shell. The density of the observed gas in the ring ranges from $(6.0 \pm 1.0) \times 10^3$ cm$^{-3}$ to $3.3 \times 10^4$ cm$^{-3}$. The early ($t \lesssim 410$ days) UV line emission is dominated by the gas with the highest density, while the late time optical emission originates in the low-density gas. Gas of lower density than $\sim 6 \times 10^3$ cm$^{-3}$ may be present in the ring, but will not be dominate the emission until after $\sim 2000$ days. The ionized mass of the gas in the ring observed up to day 1882 is $\sim 4.5 \times 10^{-2}$ M$_\odot$. The He/H ratio is $0.25 \pm 0.05$, and the overall abundance of C, N and O is $0.30 \pm 0.05$ times solar with relative abundances N/C $= 5.0 \pm 2.0$ and N/O $= 1.1 \pm 0.4$. Good fits to the light curves are obtained for a peak effective temperature of the burst in the range $(5-8) \times 10^5$ K. The corresponding color temperature is $(1.0 - 1.5) \times 10^6$ K. To model the N V $\lambda$1240 light curve, resonance scattering in a medium external to the ring is needed. It is shown that the N V line scattering can be used to discriminate between models for the formation of the inner ring and the nebulosity connected to it. In particular, the structure inferred from the N V line scattering is qualitatively consistent with that in the interacting winds model.

---

[1]Email: peter@astro.su.se

[2]Email: claes@astro.su.se

*Subject headings:* stars: circumstellar shells — stars: supernovae — stars: individual (SN 1987A) — ultraviolet: general

## 1. Introduction

Spectra of supernova (SN) 1987A have shown many emission lines of widths $\sim 10$ km s$^{-1}$ (Wampler, Richichi, & Baade 1989, Meikle et al. 1991, Cumming et al. 1996, Sonneborn et al. 1996). During the first $\sim 400$ days, observations using the International Ultraviolet Observer (*IUE*) revealed notably strong flux in He II $\lambda 1640$, C III] $\lambda 1909$, O III] $\lambda 1665$, N III] $\lambda 1751$, N IV] $\lambda 1486$ and N V $\lambda 1240$ (Fransson et al. 1989). After $\sim 400$ days, the fluxes decayed rapidly, and most lines had fallen to around noise level by day 1500 (Sonneborn et al. 1996). Subsequent observations using the Hubble Space Telescope (*HST*) showed that there was a residual flux in the He II, C III] and N III] lines up to day 2876 (Lundqvist et al. 1996), along with emission in other, weaker UV lines not detected by *IUE*. Although *IUE* has provided detailed light curves, its spatial resolution (a few arcsecs) was too low to reveal any structure of the observed gas. This contrasts the situation in the optical where there are few published fluxes (Wampler et al. 1989; Khan & Duerbeck 1991; Menzies 1991; Wang 1992; Cumming 1994), but detailed morphological information from the New Technology Telescope (*NTT*) (Wampler et al. 1990; Wang & Wampler 1992) and the *HST* (Jakobsen et al. 1991; Panagia et al. 1991; Jakobsen et al. 1994; Burrows et al. 1995; Plait et al. 1995), especially in [O III] $\lambda 5007$, but also H$\alpha$, H$\beta$ and [N II] $\lambda 6583$.

The *HST* images show that after $\sim 1000$ days the brightest narrow line emission originates in an elliptical nebula surrounding the supernova. The lack of emission inside the nebula led Jakobsen et al. (1991) to propose that the observed structure is physical, and not just a limb brightening effect. The lengths of the ellipse's semimajor and semiminor axes are $0\rlap{.}''830$ and $0\rlap{.}''605$, respectively, corresponding to a tilt angle of $\sim 43°$, if the ellipse is a projected circular ring. The thickness of the ring is $10-15\%$ of its radius (Plait et al. 1995). The tilted ring model is supported by the fact that there is a much larger velocity shear along the ellipse's minor axis than along its major (Wood & Faulkner 1989; Crotts & Heathcote 1991; Cumming et al. 1996). Cumming et al. argued that this is due to an expansion (or contraction) velocity of the ring of $10.3 \pm 0.4$ km s$^{-1}$. At the largest extension of the ellipse, the *NTT* images show that the ellipse is connected to an outer structure, extending $\sim 2\rlap{.}''5$ to the north and $\sim 2\rlap{.}''0$ to the south. Observations using the *HST* (Jakobsen et al. 1994; Burrows et al. 1995; Plait et al. 1995) revealed that the loops form two separate rings, one centered north and the other south of the supernova. In [O III] $\lambda$5007, the emission from these rings contribute $\lesssim 10\%$ to the total flux on day 1036, and it seems as if this contribution remained roughly constant thereafter (J. Wampler & L. Wang, private communication 1992). Crotts, Kunkel, & Heathcote (1995) showed that the region between rings was bright in continuum emission, presumably due to dust reflection of the light from the supernova. Later, *HST* observations showed that this region also had weak emission lines (Burrows et al. 1995). There have been reports of distinct blobs in the circumstellar medium (CSM) of the supernova, mainly observed in H$\alpha$ (Crotts, Kunkel, & McCarthy 1989; Hanuschik 1991; Cumming & Meikle 1993), but also in helium lines (Allen, Meikle, & Spyromilio 1989; Elias et al. 1993). These observations had poor spatial resolution, indicating that the blobs, perhaps with the exception of the Cumming & Meikle blob, in fact were parts of the inner and outer rings.

Fransson & Lundqvist (1989) and Lundqvist & Fransson (1991, henceforth LF91) modeled the evolution of the narrow emission lines assuming they originated in a spherical shell around the supernova. The gas was thought to be excited by the supernova outburst, the lines emerging during the subsequent phase of cooling and recombination. The model was able to account for the temporal behavior of the line fluxes during the first $\sim 500$ days, provided that the peak effective temperature, $T_{peak}$, was in the range $(3-6) \times 10^5$ K, and that the electron density was $(2-3) \times 10^4$ cm$^{-3}$. An additional, less dense, component was needed to account for the forbidden lines 700 – 1000 days after the explosion. The model agreed with the elemental abundances derived by Fransson et al. (1989) to within a factor of $\sim 2$, and provided good fits to the temperature estimates and velocities derived by Wampler et al. (1990).

However, the model had problems in explaining the fact that all UV lines started to increase *simultaneously* around days 60 – 80, as well as the observed velocity shear depending on position angle. In addition, the low velocities inferred from the lines by Crotts & Heathcote (1991) and Cumming et al. (1996) are hard to reconcile with a model in which the shell is a result of interaction between the progenitor's blue supergiant (BSG) wind with a previously lost spherically symmetric red supergiant (RSG) wind; Chevalier (1987) estimated much higher velocities for such an interaction. In this paper we show that a ring geometry, combined with detailed photoionization calculations, solves these problems in a natural way.

Lundqvist (1991) and Luo (1991) were the first to calculate the light curves for a ring geometry, and they both found $(2-3) \times 10^4$ cm$^{-3}$ for the electron density of the component emitting the bulk of the line emission. While Luo tied his model into a hydrodynamical calculation by Luo & McCray (1991), and naturally found that emission is expected also from low density components, Lundqvist assumed constant density gas. The main difference between the two studies was that while Lundqvist found a similar temperature to LF91, Luo favored a softer burst, $T_{peak} \sim 2 \times 10^5$ K. Neither Lundqvist, nor Luo, could explain the N V $\lambda1240$ light curve. Later, Blondin & Lundqvist (1993) showed that the hydrodynamical method on which Luo's analysis was based is unlikely to provide an accurate description for the formation of the ring. The more detailed calculations of Blondin & Lundqvist have recently been confirmed by Martin & Arnett (1995). This should affect somewhat Luo's conclusions about the properties of the ring and the outburst. Here we use a different approach from Luo. Instead of photoionizing a precalculated strucure of the CSM, we try to *derive* a structure from the light curves, and the *NTT* and *HST* images. We then compare our results to hydrodynamical simulations for the formation of the ring. These can be divided into two main categories: the interacting winds model in which a spherically symmetric BSG wind is overtaking an asymmetric RSG wind (Luo & McCray 1991; Wang & Mazzali 1992; Blondin & Lundqvist 1993; Martin & Arnett 1995), and the colliding winds model where the wind from the progenitor is interacting with the wind from a companion star (e.g., Lloyd, O'Brien, & Kahn 1995). Extensions to the interacting winds model include magnetohydrodynamic effects (Chevalier & Luo 1994) and photoionization of the wind by the progenitor (Chevalier & Dwarkadas 1995). There is also the model by McCray & Lin (1994) in which the inner ring is the ionized part of a protostellar disk.

In addition to the comparison to hydrodynamic models, we perform a more detailed abundance analysis than LF91 and also discuss the ionizing burst in some detail. We concentrate on modeling the UV lines observed by *IUE*. In a subsequent paper (Lundqvist et al. 1996) we will discuss the *HST* and ground-based observations in greater detail. The general outline of the paper is as follows: in §2 we describe our model, and in §3 we present our results. In §4 we discuss some implications of these, and in §5 we summarize our conclusions. Preliminary results can be found in Lundqvist (1994).

## 2. Model

We assume the ring to be circular (Gould 1994; Lundqvist 1994, Sonneborn et al. 1996) with an inner radius of $6.3 \times 10^{17}$ cm, and inclined to the line of sight by 45° (e.g., Plait et al. 1995; Sonneborn et al. 1996). The ring is ionized by the supernova outburst, and then left to recombine and cool. Our approach is similar to that by LF91, except for the geometry and our aim to model the observed light curves in greater detail. Also, we have updated much of our atomic data, modified our time dependent photoionization/recombination code to better handle optically thick situations, as well as included more recent calculations for the radiation from the outburst. We now discuss these refinements.

## 2.1. The Outburst

LF91 and Luo (1991) used the results from two hydrodynamical calculations for the outburst: the 10L model by Woosley (1988) and the 11E1Y6 model by Shigeyama, Nomoto, & Hashimoto (1988). The values of $T_{peak}$ in these models are $\sim 2 \times 10^5$ K and $\sim 4 \times 10^5$ K, respectively. The full evolution of the bursts were included in both LF91 and Luo (1991). It was assumed in these studies that the color temperature, $T_{col}$, is equal to the effective temperature, $T_{eff}$. With this simplification, LF91 found that for an optically thin, spherical shell, $T_{peak} = (3 - 6) \times 10^5$ K gave the best agreement with the observations, whereas Luo (1991) argued that the 10L model provides a better fit to the burst.

Later, Shigeyama & Nomoto (1990) refined their calculations, using a finer zoning for the outer supernova envelope than Shigeyama et al. (1989). With this improvement they found for a model similar to 11E1Y6, designated 11E0.8, $T_{peak} \approx 5.5 \times 10^5$ K instead of $T_{peak} \approx 4.0 \times 10^5$ K. Ensman & Burrows (1992) made further refinements, allowing for a difference between the gas and radiation temperatures, as well as the formation of a viscous shock a few minutes after the start of the outburst. In one of their models, called 500full1, they too obtained $T_{peak} \approx 5.5 \times 10^5$ K. The general evolution of $T_{eff}$ in this model resembles closely that of 11E0.8, though the bolometric luminosity is slightly higher in 500full1 than in 11E0.8. The other model calculated by Ensman & Burrows, called 500full2, has a somewhat higher $T_{peak} \approx 7.3 \times 10^5$ K, as well as higher bolometric luminosity and number of ionizing photons. This is because the explosion energy in 500full2 ($\sim 2 \times 10^{51}$ erg) is roughly twice as high as in 500full1. Calculations using a fine spatial zoning were also made by Blinnikov & Nadyozhin (1991). Though they treated Compton scattering as pure absorption they obtained temperature and density structures very similar to those of Ensman & Burrows. The effective temperature in the models of Blinnikov & Nadyozhin is close to that in the 11E0.8 and 500full1 models, $T_{peak} \sim 5 \times 10^5$ K.

Ensman & Burrows (1992) distinguished between $T_{eff}$ and $T_{col}$. In 500full1, the peak value of $T_{col}$ is $\sim 1.2 \times 10^6$ K, which is a factor of 2.2 higher than $T_{peak}$. In order to simplify the inclusion of Ensman & Burrow's models in our calculations we follow the suggestion by Lundqvist (1992) to use a diluted blackbody with the 'effective' temperature equal to $T_{col}$; the absolute flux is scaled down by a factor $q(t) = (T_{eff}/T_{col})^4$ to preserve the total luminosity. We also include the difference in light travel time of the ionizing photons emitted from the limb of the supernova compared to those emitted from the projected center of the supernova. This causes a spread of the burst by $\sim R_p/c$, where $R_p$ is the photospheric radius of the supernova. For 500full1 this corresponds to $\sim 1.1 \times 10^2$ s. A larger spread in arrival time of the ionizing photons may be introduced were the explosion and/or the presupernova not spherically symmetric (Lundqvist 1992). This may lower the temperature in the inner part of the ionized ring (Lundqvist 1992) compared to calculations assuming spherical symmetry, whereas the ionization structure is less affected. As in Fransson & Lundqvist (1989) and LF91, we join the calculated values of $T_{eff}$ to that deduced from *IUE* observations 1.5 days after core collpase, $\sim 1.5 \times 10^4$ K (Cassatella et al. 1987; Kirshner et al. 1987). No attempts have been made to include limb darkening effects.

In Table 1 we list the most important properties of four burst models: 10L, 11E0.8, 500full1 and 500full2. The 10L and 11E0.8 bursts yield roughly the same number of ionizing photons, though the energy distributions of the photons differ significantly. The energy distributions in 11E0.8 and 500full1 are similar, but the total number of ionizing photons in 500full1 is a factor of $\sim 1.5$ higher than in 11E0.8. The 500full2 burst produces another factor of $\sim 2.2$ more ionizing photons. The bursts in 10L, 11E0.8, 500full1 and 500full2 are capable of ionizing 0.49, 0.58, 0.87 and 1.91 $M_\odot$, respectively, assuming the CSM is pure hydrogen. For a ring geometry these masses are scaled down by a factor of $\Omega/4\pi$, where $\Omega$ is the solid angle subtended by the ring as seen from the supernova.

Klein & Chevalier (1978) found for normal Type II SNe that a considerable amount of energy is emitted in the keV range as a result of the formation of a viscous shock. However, the existence of this is controversial, and depends on the efficiency of preacceleration of the gas (Epstein 1981). If included, the X-ray emission should be added to the otherwise roughly blackbody-shaped burst spectrum. Ensman & Burrows (1992) found that a viscous shock probably formed in SN 1987A, but that the survival and temperature of the thin shell driving the shock depend on the amount of circumstellar matter close to the star. To test whether such an X-ray tail is important for the circumstellar lines, and the N V $\lambda 1240$ line in particular as was proposed by Luo (1991), we have added a free-free tail to the blackbody component between 0.01 and 0.03 days after the start of the outburst.

### 2.2. Model Parameters

We approximate the shape of the ring by a wedge. As viewed from the supernova, the wedge has an opening angle, $\theta$ (see Fig. 6), which remains constant with radius throughout the ring. For the ionizing burst we use mainly the 500full-models by Ensman & Burrows (1992). Density, abundances, and the mass of the ionized gas are then adjusted to match the observed line fluxes.

For many of the ions we increase the number of levels for the model atoms compared to LF91, and we have also updated much of the atomic data. This is summarized in Appendix A, and a short discussion on how we have improved our numerical methods are given in Appendix B.

To model the *observed* line fluxes we include light travel time effects similar to in LF91, but since we now deal with a ring geometry instead of a spherically symmetric shell, the convolution function is different. This was given for an infinitesimally thin ring by Dwek & Felten (1992). Here we take into account the radial extent of the ring, which smooths somewhat the sharp peaks in the light curves introduced by the assumption of an infinitesimally thin ring.

We compare our results against the *IUE* observations (Sonneborn et al. 1996), and observations in the optical by Wampler et al. (1989), Khan & Duerbeck (1991), Menzies (1991), Wang (1992) and Cumming (1994). The observed fluxes are dereddened using $E_{B-V} = 0.16$, with $E_{B-V} = 0.06$ from the Galaxy and $E_{B-V} = 0.10$ from the LMC (Sonneborn et al. 1996). The extinction curves are taken from Savage & Mathis (1979) and Fitzpatrick (1985).

## 3. Results

### 3.1. Temperature and Ionization

In Figure 1 we show the results for a model with density $n = 2.0 \times 10^4$ cm$^{-3}$. With the elemental abundances described below, this corresponds to an electron density, $n_e \approx 2.4 \times 10^4$ cm$^{-3}$ when the gas is completely ionized. This value for $n_e$ is close to that derived by Fransson (1988) for the C III] $\lambda\lambda$1907, 1909 doublet (see also Sonneborn et al. 1996), and is close to the density used by LF91, Lundqvist (1991) and Luo (1991). For the outburst we use the 500full1 model, and include elements with relative abundances H : He : C : N : O : Ne : Si : S : Fe = 1 : 0.25 : $3.2 \times 10^{-5}$ : $1.8 \times 10^{-4}$ : $1.6 \times 10^{-4}$ : $5.5 \times 10^{-5}$ : $1.7 \times 10^{-5}$ : $5.6 \times 10^{-6}$ : $3.0 \times 10^{-5}$. These correspond to the total abundance of C, N and O being 0.30 times solar. The abundances of H, He, C, N and O were chosen to give good fits to the *IUE* and optical light curves of the narrow lines (see §3.3). The high helium abundance is in accord with the results by Shigeyama et al. (1988) and Allen et al. (1989). The abundances of neon and iron are from Barlow (1989) and Russell, Bessell, & Dopita (1988), respectively, and that of sulphur is close to what is suggested by Pagel et al. (1978) for the LMC. For silicon we assume that the O/Si ratio is the same in the LMC as in the Galaxy, just as it appears to be for O/S (Pagel et al. 1978). We will *derive* the abundances of He, Ne, Si, S and Fe in Lundqvist et al. (1996).

Figure 1 shows that during the ionization, the He III zone is heated to temperatures in excess of $10^5$ K, with a peak value of $\sim 1.8 \times 10^5$ K at $\sim 5 \times 10^{14}$ cm from the inner edge of the ring. The fact that the temperature increases with radius in the inner part of the ring is an optical depth effect. As the mean energy of the emitted photons from the burst decreases with time, there will be a turn-over in the radial temperature structure of the ring. Exterior to the turn-over, the decreasing temperature with radius is an imprint of the decreasing mean photon energy with time emitted by the supernova. The thermal structure is thus dependent on the exact evolution of the burst (Lundqvist 1992), while this affects the ionization structure to a smaller extent. Figure 1 shows that the ionization structure is characterized by very broad He II and O III zones, whereas higher ionization stages have rather narrow zones and occur close to the inner edge of the ring. The ionization boundary of H is at $\sim 9.5 \times 10^{15}$ cm from the inner edge. The radial thickness of the ionized part of the ring is therefore only $\sim 1.5\%$ of the ring radius.

Subsequent to the ionization, the ring cools mainly by UV and EUV lines in its inner parts (see also LF91), and by optical lines further out. Figure 1 shows that the temperature decreases with time by a factor of $\sim 2$ throughout the H II zone during the first 200 days. Further out, the temperature decrease is marginal due to the lack of free electrons, and the inefficient collisional excitation of ions due to collisions with neutral hydrogen.

During the same period, metals like C, N and O recombine to ionization stage III in the inner part of the ring, while the recombination time for H II and He III is too long to affect significantly the ionization structure of H and He. Again, the O III zone is broad, but has shifted to smaller radii compared to the structure shortly after the ionization. An even broader N II zone develops during the first 200 days. The temperature in the O III zone remains rather constant with time, while that in the N II zone decays somewhat. This trend continues up to late times (see Fig. 1). At 2000 days, N and O have recombined almost completely, and also H and He have large neutral populations. The temperature, $\sim 5 \times 10^3$ K, is at this epoch fairly constant throughout the gas.

While the initial heating and ionization is nearly independent of density, the time scale for the subsequent cooling and recombination scales as $\sim n_e^{-1}$. The evolution of the temperature and ionization structure seen in Figure 1 is therefore easy to transform to other densities; the evolutionary time scale increases by a factor of $\sim (n/2.0 \times 10^4 \text{ cm}^{-3})^{-1}$. Some forbidden lines become collisionally deexcited as the density increases, and fine-structure lines are in LTE, regardless of the density. However, this has only a marginal effect on the general $n_e^{-1}$ scaling.

To study how the ionization and temperature structure respond to a different burst spectrum, we have run another model replacing 500full1 by 500full2. The density is still $n = 2.0 \times 10^4$ cm$^{-3}$. Because the 500full2 burst has a harder spectrum and more ionizing photons than 500full1 (Table 1), the ring is hotter and more highly ionized, compared to what is shown in Figure 1. In the case of 500full2, the maximum temperature of the ring,

$\sim 2.2 \times 10^5$ K, occurs $\sim 7 \times 10^{14}$ cm away from the inner edge of the ring. Nitrogen is at the inner edge ionized up to equal populations of N VI and N VII, and oxygen is dominated by O VII, with a fraction of $\sim 10\%$ even reaching O VIII. The H II zone reaches $\sim 2.1 \times 10^{16}$ cm from the inner edge of the ring. Qualitatively, the outermost part of the ionized region is similar to that shown in Figure 1. In fact, if the distance scale in Figure 1 were compressed by the ratio of the total number of ionizing photons for the 500full2 and 500full1 bursts (see Table 1), the outermost $\sim 75\%$ of the H II zone in Figure 1 would describe well a model using 500full2 instead of 500full1.

In the case of 500full2, the highly ionized ions close to the inner edge of the ring are inefficient in cooling the gas, so the innermost $\sim 10\%$ of the H II zone is almost a factor of $\sim 2$ hotter than in Figure 1 on day 200. For the rest of H II zone at this epoch, the difference in temperature is much smaller. By day 2000 the temperature and ionization structures in the two models are similar throughout the ring.

For a full discussion on the 10L and 11E1Y6 bursts, we refer to LF91 (see also Lundqvist 1992). The major difference compared to the calculations in the current paper is that the somewhat harder 11E0.8 spectrum ionizes the gas to slightly higher ionization stages than 11E1Y6. Results using 11E0.8 are similar to what is described in Figure 1, though the H II region only reaches out to $\sim 6.5 \times 10^{15}$ cm from the inner edge of the ring, instead of $\sim 9.5 \times 10^{15}$ cm as in the case of 500full1, assuming $n = 2.0 \times 10^4$ cm$^{-3}$. The maximum temperatures in the ring using 10L or 11E0.8 are $\sim 9.2 \times 10^4$ K and $\sim 1.7 \times 10^5$ K, respectively. In the case of 10L, the gas is not ionized more than up to N V.

### 3.2. Emission Lines

In Figure 2 we show the intensity of important emission lines for the model in Figure 1, assuming a distance of 50 kpc to the supernova. In order to get the right absolute flux, the opening angle of the wedge (cf. §2.2) has been set to $\theta = 2°.5$, corresponding to $2.9 \times 10^{-2}$ M$_\odot$ for the mass of the H II region.

Figure 2 shows that a single-density model is able to model some features of the light curves rather well: both the turn on and the maximum of the UV lines occur at about the right time, and the N III] $\lambda 1751$/N IV] $\lambda 1486$ ratio is close to the observed. This can be compared with the spherical shell model in LF91 which had too broad a peak and too late a turn on of the N III] line.

The most notable discrepancy in Figure 2 is for N V $\lambda 1240$; the modeled light curve represents only a marginal improvement compared to the results in LF91 in that Figure 2 shows the right turn on of the line. As noted in §3.1, electron density is an important parameter for the evolution of the light curves. A better fit for the N V line than in Figure 2 is shown in Figure 3, where $n = 2.0 \times 10^3$ cm$^{-3}$ and $\theta = 5°\!.4$, giving an ionized mass of $5.6 \times 10^{-2}$ M$_\odot$. Abundances are the same as in Figure 2. However, such a low density ruins the fit for other lines; in particular, the C III], N III] and N IV] light curves have too prominent tails due to long recombination times from ionization stages V and VI, caused by the low density. A low density can therefore not be the sole explanation to the N V light curve. A more likely explanation is that non-local resonance scattering transfers flux from the early part of the light curve ($\lesssim 400$ days) to the tail, because of light travel time effects. We return to this in §3.5 (see also Lundqvist 1994). Here we note that the time integrated N V $\lambda 1240$ flux in Figure 2 is only slightly lower than the observed.

Another feature which is not modeled well in Figure 2 is the tail of the light curves for the O III], N III] and N IV] lines, indicating that recombination is too rapid in this model. This is most likely a result of our assumption of only one density component; the tails are probably due to a low density component. The fact that more than one density component is required is not surprising, considering the clumpy structure of the ring (e.g., Plait et al. 1995). A multi-density structure is also a natural outcome of hydrodynamical simulations for the formation of the ring (Luo 1991; Blondin & Lundqvist 1993). We return to this in §3.3.

In Figure 2 we assumed He/H = 0.25. This is a preliminary value to the results of Lundqvist et al. (1996), where the optical lines are modeled in greater detail. As discussed in that paper, the He/H-ratio is sensitive to the rather uncertain collision strengths for hydrogen for transitions from $n = 1$ to $n = 3$, and especially from $n = 1$ to $n = 4$ (see Appendix A). If the Balmer lines were only powered by recombination, the absolute flux level of these lines would be roughly half the observed during the first $\sim 800$ days. This shows that an abundance analysis based on recombination only, can give an incorrectly low He/H-ratio. For the model in Figure 2, collisional excitation is important for the first $\sim 1500$ days. This means that the inclusion of collisional excitation is essential in order to derive reliable He/H-ratios during the first $\sim 1500 \, (n/2.0 \times 10^4 \text{ cm}^{-3})^{-1}$ days. Because the density of the gas dominating the Balmer line emission tends to decrease with time (Lundqvist et al. 1996), collisional excitation is always important.

In order to study the sensitivity of the line emission to bursts different from 500full1, we show in Figure 4 results for a model similar to that in Figure 2, this time using the 500full2 burst. Again, the density is $n = 2.0 \times 10^4$ cm$^{-3}$. Because the number of ionizing photons is larger than for the 500full1 burst, the opening angle $\theta = 1°\!.3$ is only half as large as in Figure 2, though the mass of the H II region is nearly the same, $3.1 \times 10^{-2}$ M$_\odot$.

Despite the differences between 500full1 and 500full2, Figures 2 and 4 are remarkably similar, especially the light curves of the optical lines. As discussed in §3.1; the outermost 75% of the H II region in the two models are almost identical. Because this is from where a large portion of the optical line emission originates, we conclude that *the optical lines are rather insensitive to the burst spectrum; the main parameter determining their light curves is the gas density*. The same conclusion is made by Cumming et al. (1996) when they model the line profiles of the optical lines. The only real difference between the models in Figures 2 and 4 occurs for the region close to the inner edge of the ring. Since this is where the UV lines form, it is only these that show a different behavior. In particular, this applies to the N V line, which has an integrated luminosity closer to the observed in Figure 4 compared to Figure 2. This is because the temperature of the N V zone is higher in the case of 500full2, boosting the collisional excitation of N V. The *shape* of the N V $\lambda 1240$ light curve is not modeled well using any of the burst models, and the reason for this is discussed in §3.5. Another difference between the models is that the carbon abundance had to be increased in the model in Figure 4 so that N/C = 4.5, instead of N/C = 5.5 as in Figure 2. All other metal abundances are the same to within a few per cent.

The trend described above that the UV lines, in particular the N V line, are sensitive to the burst spectrum, while the optical lines are not, is seen also for models with lower values of $T_{peak}$ than in 500full1. We find that the 10L model can be ruled out, because it produces far too low a time integrated N V $\lambda 1240$ flux, and also because it cannot model well the shape of the He II $\lambda 1640$ light curve. He II $\lambda 1640$ is dominated by collisional excitation (LF91), which is not the case if the ring is ionized by a burst like 10L. Also the spectrum in the 11E0.8 model is too soft. One way of obtaining better fits to the UV lines for the 11E0.8 and 10L models could be to invoke an X-ray tail (Luo 1991), like the one found by Klein & Chevalier (1978) in simulations of supernova outbursts (§2.1). We find that in order to do so for 10L, the cut-off energy of the tail has to be as low as $kT_x \lesssim 1$ keV, if the energy of the tail is not to exceed 50% of the total ionizing energy in 10L. This is very different from the results by Klein & Chevalier (1978) who find $kT_x \sim 7.3$ keV and $\sim 1\%$, respectively. In addition to this, X-ray tail models tend to generate too much N IV] $\lambda 1486$ compared to N III] $\lambda 1751$ and N V $\lambda 1240$. Therefore, we find it unlikely that the N V $\lambda 1240$ light curve can be explained by an X-ray tail. In §3.5 we discuss a more likely explanation based on non-local resonance scattering.

We have also run models invoking asymmetry of the explosion, in an approximate way, by stretching out the burst in time by up to five times longer than is introduced by the light travel time effect discussed in §2.1. Our results show that changing the characteristic of the burst this way only marginally affects the shape of the light curves.

### 3.3. Mixed density model: Densities and abundances

From §3.2 it is evident that one cannot explain the shape of all light curves using a single-density model for the ring. In a multi-density model, the density components can be arranged geometrically in many ways; the components may, or may not, block each other from the ionizing burst. The two limiting cases are that all components are exposed directly to the burst (henceforth called Case 1), or alternatively, are arranged radially in a stratified manner so that the outer components see the burst through the inner components (Case 2). Though the bulk of the UV line emission at early times is dominated by high-density gas, the late emission requires $n \lesssim 2.0 \times 10^4$ cm$^{-3}$ (§3.2). This immediately tells us that both low- and high-density gas must coexist close to the inner edge of the ring, since this is where the UV lines are formed. In Case 1 this is not a problem, but in Case 2, the innermost zone must be thin to allow for an abrupt change in density within the UV line emitting region.

It is not possible to make a fully unique model when different density components are combined; the number of components are not known, and the combination of components depends to some extent also on the characteristics of the burst. We will limit ourselves to the minimum number of components that are necessary to explain adequately the evolution of the line emission. The idea of having components with a few discrete densities is, of course, a simplification; it is more likely that a continuous range of densities is present.

In Figure 5 we show a Case 1 model for the 500full1 burst with three components: $n = 3.3 \times 10^4$ cm$^{-3}$, $n = 1.4 \times 10^4$ cm$^{-3}$ and $n = 5 \times 10^3$ cm$^{-3}$. The masses of the H II-regions of the components are $1.2 \times 10^{-2}$ M$_\odot$, $1.0 \times 10^{-2}$ M$_\odot$ and $2.4 \times 10^{-2}$ M$_\odot$, respectively, i.e., the total mass is $\sim 60\%$ higher than in Figure 2. The combination of components in Figure 5 were chosen to model both the lines discussed in this paper, and the optical lines discussed in Lundqvist et al. (1996) up to around day 2000. To fit the light curves we find that the elemental abundances are insensitive to the exact combination of components; in Figure 5 the abundances are the same as in Figure 2. The observationally well-defined UV line light curves enable us to estimate the C : N : O and O/He ratios, and we find 1 : 5.5 : 5.0 and $\sim 6.5 \times 10^{-4}$, respectively. Preliminary results to the analysis by Lundqvist et al. (1996) of *HST* and ground based spectra give a value for the He/H-ratio of $0.25 \pm 0.05$. The uncertainty in the hydrogen abundance implies a similar uncertainty ($\pm 20\%$) in the abundance of C+N+O around the value 0.30 times solar (Lundqvist et al. 1996). In Figure 5 we have included the resonance scattering of the N V line as discussed in §3.5. We note that the integrated flux is closer to the observed than in the single-density model in Figure 2, but that it is still somewhat smaller than the observed.

Using the 500full2 burst we obtain good fits to the light curves for a combination of components with densities $n = 3.3 \times 10^4$ cm$^{-3}$, $n = 1.4 \times 10^4$ cm$^{-3}$ and $n = 7.0 \times 10^3$ cm$^{-3}$.

These densities are nearly identical to those in the model displayed in Figure 5 for 500full1. The masses of the H II-regions of the components are $1.3 \times 10^{-2}$ $M_\odot$, $7.3 \times 10^{-3}$ $M_\odot$ and $2.2 \times 10^{-2}$ $M_\odot$, respectively, the total mass being close to that in Figure 4. For C, N and O we now obtain C : N : O = 1.0 : 4.5 : 4.0, while the O/He ratio is $\sim 6.6 \times 10^{-4}$, i.e., close to that for the model using 500full1. The overall metallicity and He/H ratio are the same as for 500full1. The main difference between the calculations using 500full1 and 500full2 is the flux level of the N V line (see Fig. 7). It is slightly higher, and closer to the observed flux, in the case of 500full2.

It is also possible to obtain reasonable fits for models where the components obscure each other (i.e., Case 2 models), using either of the two 500full bursts. The best Case 2 fits have a high density component facing the supernova, and a less dense component located exterior to this. For example, for 500full1, the densities of the inner and outer components are $n = 5.0 \times 10^4$ cm$^{-3}$ and $n = 4.0 \times 10^3$ cm$^{-3}$, respectively. The interface between these components should be located within the UV line emitting region to allow for UV line emission from both components (cf. above). We obtain our best fits when the interface is located $\sim 15\%$ from the inner edge of the H II region. The total mass of the H II region needed is $\approx 4.5 \times 10^{-2}$ $M_\odot$, corresponding to $\theta \approx 4°\!.0$. Abundances are the same as in all 500full1 models discussed above, which leads us to the conclusion that *the elemental abundances are insensitive to the geometry.* We cannot exclude Case 2 models on the basis of the UV line light curves alone. In fact, the integrated N V line flux is closer to the observed than is shown in Figure 5. Possible problems with Case 2 models are rather broad spikes in the He II and N IV] light curves around day 400, as well as somewhat too high an N IV]/N III] line ratio. The [N II] and [O III] lines are both underproduced by a factor of $\sim 1.5$ at epochs earlier than $\sim 1000$ days. To some extent, this can be avoided by mixing high-density gas into the low-density component. The situation will thus be increasingly similar to Case 1, the more high-density gas is mixed into the low-density component. We therefore believe that Case 1 models are closer to the real structure than a pure Case 2 structure.

In the Case 1 models we find that the lowest density required by the data is $n = (6.0 \pm 1.0) \times 10^3$ cm$^{-3}$, and that the mass of ionized gas is $\sim 4.5 \times 10^{-2}$ $M_\odot$. However, there may be substantial amounts of gas of lower density which will increase the ionized mass. Such low-density gas will not dominate the emission until after $\sim 2000$ days. A discussion on the low-density gas is provided by Lundqvist et al. (1996).

### 3.4. Burst Characteristics

As discussed above, both 500full1 and 500full2 can be used to model the light curves to within the observational uncertainties. The time integrated flux of the N V line in models using the 500full1 burst is on the low side, but is within the uncertainties. The 11E0.8 model has the same value of $T_{peak}$ as 500full1, but can be excluded because this burst results in too low a N V $\lambda1240$ flux. The reasons for this are that 11E0.8 has a somewhat lower overall energy ouput than 500full1, and that it inherently assumes that $T_{col} = T_{eff}$. When $T_{col}$ is allowed to deviate from $T_{eff}$, as in the 500full models, the lower limit to $T_{peak}$ is $\sim 5 \times 10^5$ K, and the number of ionizing photons has to be in excess of $\sim 10^{57}$ to explain the flux level of N V $\lambda1240$. The upper limit on $T_{peak}$ depends on the ratio of $T_{col}/T_{eff}$. For $T_{col}/T_{eff} \approx 2$, as in 500full2, $T_{peak}$ cannot be much higher than in 500full2 in order not to overproduce the N IV] and N V lines, resulting in an upper limit $T_{peak} \sim 8 \times 10^5$ K. To obtain a higher effective temperature, using the same model for the progenitor as in 500full1 and 500full2, Ensman & Burrows (1992) would need an explosion energy reaching $3 \times 10^{51}$ erg. The conclusions in Fransson & Lundqvist (1989) about $T_{peak}$ are thus unaltered. The fact that 11E1Y6 (similar to 11E0.8) was found to be sufficient in LF91 was because of a rather crude procedure to fit the evolution of the burst. This procedure overestimated the number of ionizing photons so that 11E1Y6 appeared to be energetic enough to power the narrow UV lines. Luo (1991) suggested that it is necessary to invoke a substantial X-ray tail to explain the N V line. As discussed in §3.3, we do not find this necessary; using the 500full1 or 500full2 bursts is sufficient to model the observed N V $\lambda1240$ flux.

### 3.5. The N V $\lambda1240$ Light Curve

In §3.2 we noted that the persistence of the N V $\lambda1240$ flux cannot simply be attributed to a low density component. Such a component would give rise to an even longer duration of the other UV lines, which, in particular for N IV] $\lambda1486$, is in strong conflict with the observations (see Fig. 2). We must therefore seek another explanation for the late N V $\lambda1240$ emission. Noting that the N V $\lambda1240$ flux in the models in Figures 5 and 7 is too high at early times and too low at late times, while the *time integrated* flux is modeled well, our problem is not to reproduce the total flux, but to redistribute the emission from early to late times. Because N V $\lambda1240$ is the only resonance line of the observed lines, the natural choice of redistribution is resonance scattering.

For resonance scattering to be successful, the photons should be trapped for a few hundred days. This puts limits to the velocity and density structure of the scattering gas; N V $\lambda1240$ can only be scattered as long as there is N V present in the gas, and the scattering and emitting regions have a relative velocity less than the width of the line. The scattering gas also has to subtend a large enough solid angle, as seen by the emitting gas, to have an observable effect. An obvious possibility is that the scattering gas is mixed with, or even identical to the emitting gas in the ring. However, this scenario fails because of the limited size of the scattering region. The size of the scattering region, $\Delta$, is in this case equal to the thickness of the observed ring, $\Delta \sim 10^{17}$ cm (e.g., Plait et al. 1995), giving a time for a photon to escape the scattering gas of only $t_{esc} \sim \Delta/c \sim 39$ days.

To increase $t_{esc}$, the size of the scattering region has to be much larger than the observed thickness of the ring. Because scattering from one side of the inner ring to the other is unimportant due to small solid angles and the velocity shift, we have to appeal to an external medium which has negligible contribution to the total line *emission*, both in N V $\lambda$1240 and other lines. The size of the scattering region should be of order $\Delta \sim 10^{18}$ cm to give the right duration of the N V $\lambda$1240 flux. Furthermore, the scattering gas has to be close to the emitting gas to subtend a large enough solid angle, and the velocity difference between the emitting and scattering regions must be of order the thermal velocity of the scattering gas. Direct evidence for gas extending from the inner ring, possibly extending to the outer rings, comes both from ground-based observations (e.g., Crotts et al. 1995) and observations using the *HST* (Burrows et al. 1995). The interacting winds model (henceforth the IW model) as developed by Blondin & Lundqvist (1993) and Martin & Arnett (1995), and the colliding winds model (henceforth the CW model) developed by Lloyd et al. (1995), both provide two possible candidates for such a low density, extended gas: the interaction region between the BSG and RSG winds, and the unshocked RSG wind. Whenever we use the abbreviations IW or CW we refer to the model calculations by these three groups.

In the IW model, the shocked RSG wind curves outwards from the inner ring, where it has its smallest distance to the supernova (cf. Fig. 6). The calculations show that the thickness and shape, as well as density and velocity structures of this gas, depend primarily on the degree of asymmetry of the RSG wind, but also on the velocity field of the wind and the efficiency of radiative cooling. The shape of the nebula in the case of the CW model has not been demonstrated in detail, but is likely to be qualitatively similar to the IW model, though perhaps more complex.

To test our hypothesis of resonance scattering, we discuss two structures for the scattering gas: first, a geometrically simple model to better reveal the physics involved, and second, the model h of Blondin & Lundqvist (1993; their figure 8; henceforth BL8 and shown in Fig. 6). In the former case we assume that the scattering gas is confined to a cylinder with inner radius slightly larger than the ring radius. This is a rough approximation to the the IW model, and presumably also to the CW model. In the cylinder model we neglect recombination of the scattering gas and assume that the density is constant within the cylinder. This approximation is valid if the density is $\lesssim$ few $\times\,10^2$ cm$^{-3}$. The importance of recombination is discussed below for BL8.

The parameters of the cylinder model are thus the thickness, $\Delta_c$, and height, $h_c$, of the cylinder, the N V column density through the cylinder in the plane of the ring, the temperature, $T_c$, of the scattering gas, and the velocity, $V_c$, with which the cylinder recedes from the supernova in the plane of the ring. The optical depth in a given direction at the line center of the emitted photons is then given by $\tau_c(\nu_o) = \int \kappa_o(l)\,\phi[T(l), \nu_o\,(1+V(l)/c)]\,dl$, where $\kappa_o$ is the opacity at the line center, $\phi$ the line profile function, here taken to be the Voigt profile, and $V(l)$ the recession velocity of the scattering gas relative to the emitting gas. The optical depth $\tau_c$ therefore depends not only on the N V distribution and the temperature structure, but also on the velocity field. In the absence of a velocity gradient $\tau_{lc} = \int \kappa_o(l)\,dl$. Obviously, $\tau_c < \tau_{lc}$. We now discuss the constraints on these parameters. Although we may put limits on these from observations and model calculations, we caution that it is not possible to derive a unique set of parameters. Detailed hydrodynamical simulations are therefore necessary to provide enough constraints.

As was argued above, the size of the region should be $h_c \sim 10^{18}$ cm. In addition, the scattering gas should be close to the ring to give a large solid angle. A reasonable fit to BL8 is that the radius of the cylinder is in the range $(1.3 - 1.5) \times R_s$. The expansion velocity for the various parts of the scattering gas is assumed to scale with the distance from the supernova. This is a good approximation to the results in the IW model (Blondin & Lundqvist 1993). Furthermore, $\tau_c$ along the line of sight to the near side of the ring cannot be much greater than unity. Otherwise, the initial rise of the line flux would be blocked out completely.

To see how the temperature and fraction of N V vary with distance from the supernova, we have used our photoionization code to calculate the ionization structure of a gas element placed at distances ranging from $6.3 \times 10^{17}$ cm ($= R_s$) to $2.5 \times 10^{18}$ cm. We use the same elemental abundances and models for the burst as in §3.3. The gas is assumed to be ionization bounded, i.e., much thicker than the N V zone. This is also a reasonable approximation to the results of Blondin & Lundqvist (see below).

The values of $\tau_{lc}$ and $T_c$ obtained from our photoionization calculations are summarized in Table 2 for $n = 2.0 \times 10^4$ cm$^{-3}$. The value of $T_c$ is the temperature where the fraction of N V is the highest. Results are given for 1, 10 and 100 days. For other densities one obtains similar temperatures and optical depths at an epoch $\sim t\,(n/2.0 \times 10^4$ cm$^{-3})-1$. Both $\tau_{lc}$, which reflects the fraction of the gas ionized to N V, and $T_c$ decrease with increasing distance to the supernova due to the dilution of the ionizing radiation. For the distances typical to where we find the N V scattering to take place, $T_c \approx (1.2 - 1.6) \times 10^5$ K.

The calculations above constrain $T_c$ and the column density of N V as a function of distance from the supernova. The resulting optical depth $\tau_{lc}$ translates into the velocity dependent optical depth $\tau_c$ when the velocity field is included. From the observations we know that $\tau_c$ should be of order unity, implying that $\tau_{lc} \gtrsim 1$, which is confirmed by Table 2. Guided by these constraints we adjust $\tau_{lc}$, $h_c$ and $\Delta_c$ to fit the N V light curve, taking into account that the typical length scale should be $\sim 10^{18}$ cm. The calculations for the resonance scattering of the N V line were performed using a Monte Carlo method, assuming the line has a Voigt profile in the rest frame of the scattering gas. In the cylinder model we neglect the fact that N V $\lambda\lambda 1239$, 1243 is a multiplet. The maximum velocity shift between emission and scattering is $\lesssim 10^2$ km s$^{-1}$, corresponding to $\lesssim 0.4$ Å, which means that blending of the doublet components is unimportant. However, for a detailed calculation, like for the BL8 model discussed below, the line components must be treated separately since the components have different opacities.

In Figures 5 and 7, we show results for $\Delta_c = 3 \times 10^{17}$ cm, $h_c = 10^{18}$ cm, $\tau_{lc} = 2$ and $T_c = 1.5 \times 10^5$ K. The inner radius of the cylinder is 1.3 times that of the ring, giving $V_c = 1.3 \times 10.3 \approx 13.4$ km s$^{-1}$. The Doppler width of the line for $1.5 \times 10^5$ K corresponds to 13.3 km s$^{-1}$, so the scattering optical depth through the cylinder does not fall below unity until the velocity difference between the emitting and absorbing gas is $\gtrsim 17$ km s$^{-1}$. This means that photons scatter in the part of the cylinder which is close to the part of the ring from where they originate, while they escape almost freely through distant parts of the cylinder.

Our cylinder models indicate that $h_c \sim 10^{18}$ cm is needed. Models with $h_c = 5 \times 10^{17}$ cm give too short a duration of the line flux, while for $h_c \gtrsim 2.0 \times 10^{18}$ cm the emission from the ring is trapped too long by the scattering gas. It should be emphasized, however, that a value of $h_c = 10^{18}$ cm does not necessarily mean that this is the actual size of the external gas, only the size of the region containing N V which recedes from the ring by $\lesssim 20$ km s$^{-1}$. The external gas may extend to higher latitudes, as long as it is transparent in N V $\lambda 1240$ emitted from the ring.

Table 2 shows that the line optical depth is much larger than in the best fit cylinder model ($\tau_{lc} = 2$). In particular, along the line of sight to the nearest point of the ring, the optical depth for a gas structure similar to that in the cylinder model in Figures 5 and 7 is $\sim 100$. This problem is most likely a result of our over-simplified gas distribution.

The models of Blondin & Lundqvist (1993) show that the bipolar nebula curves outwards to larger radii, as one goes from the equatorial plane to higher latitudes (see Fig. 6). In BL8 the distance along the line of sight from the nearest point of the ring to the intersection of the line of sight with the nebula is $\sim 1.8 \times R_s \approx 1.1 \times 10^{18}$ cm, and the distance from the supernova to this intersection is $\sim 2.6 \times R_s \approx 1.7 \times 10^{18}$ cm. Table 2 shows that the temperature and optical depth at this location are $(7.5 - 9.0) \times 10^4$ K and $8 - 40$, respectively. The lower values are for 500full1, and the upper for 500full2. The velocity shift between the emitting and scattering gas is in BL8 $\sim 19.5$ km s$^{-1}$, corresponding to 2.0 Doppler widths for a gas temperature of $8.0 \times 10^4$ K. The value of $\tau_{lc}$ along the line of sight to the near side of the ring is thus $\lesssim 1.0$, and may actually be too low to give a good fit to the the early part of the N V $\lambda$1240 light curve.

To test the BL8 model in greater detail, we have used our time dependent photoionization code to follow the ionization and recombination of the scattering gas. The BL8 nebula is ionized using the 500full bursts and assuming the abundances derived in §3.3. The structure of the N V-rich gas shortly after ionization is shown in Figure 6 for the 500full1 case. The nebula is then left to recombine and cool according to its density structure. To calculate the scattering of the N V $\lambda$1240 photons we used a Monte Carlo method. The photons are released only in the equatorial plane of the recombining BL8 nebula (i.e., the inner ring), and the starting point of the photons is selected according to the radial distribution of the N V $\lambda$1240 emissivity. The photons then experience multiple scattering in the nebula (Fig. 6), and from this we obtain a distribution function for the time delay of the photons directed to the Earth. Photons coming from the near side of the ring are more likely to be affected by scattering (Fig. 6; path $B$ to $D$), than those from the far side (Fig. 6; point $F$). The reason to this is the large velocity difference between far side of the ring (point $F$) and the near side of the nebula along the line of sight. The distribution function for the time delay is convolved with the evolution of the N V $\lambda$1240 instantaneous emission from the ring in the three-component models in §3.3 to compute modeled light curves. The instantaneous emission is the emission from the ring with no light echo effects included. Because the density distribution in BL8 is most likely different from the distribution in the observed ring with its multi-density structure (§3.3), local scattering in the ring in our calculations is inaccurate on a time scale $\lesssim 40$ days. The two components of the line are treated separately since the optical depth in N V $\lambda$1243 is half that in N V $\lambda$1239, and the lines therefore have different time delays. Both the absorption and emission line profiles are assumed to have Voigt profiles, and the photons are assumed to be completely redistributed in direction and frequency at each scattering event. The full 2-D velocity and density structures of BL8 are included, using cylindrical symmetry to extend these to 3-D. The nebula is tilted by 45° to the line of sight.

The resulting light curves are shown in Figure 7, both for 500full1 and 500full2. To study the sensitivity to the density we have in one set of the calculations assumed that no recombination occurs (corresponding to $n_e \lesssim 10^2$ cm$^{-3}$), and in the other (lower panel of Figure 7) included recombination as given by the density in BL8. The 500full2 burst is capable of ionizing the gas to N V throughout the nebula, whereas in the case of 500full1, the gas is deficient in N V at large latitudes. Large scale scattering, as exemplified from $B$ to $D$ in Figure 6, is therefore more important for a burst like 500full2, resulting in a pronounced tail of the N V $\lambda 1240$ light curve. Because the gas at large latitudes has a low density (of order $\sim 10^2$ cm$^{-3}$), the large scale scattering, and thus the tail of the N V $\lambda 1240$ light curve, are unaffected by recombination, as can be seen in Figure 7. Closer to the equatorial plane, where the density is higher, recombination is more important, in particular for 500full1; for 500full2, the gas at low latitudes is ionized to N VI, which recombines only slowly to N V, whereas for 500full1, the gas is less ionized, and therefore becomes deficient in N V at a faster rate. As can be seen in Figure 7, recombination boosts the early part of the light curve compared to the non-recombining case.

From Figure 7 it is evident that the model light curves for BL8 have too low a flux between $\sim 200$ and $\sim 400$ days. This is due to the large curvature outwards from the equatorial plane in BL8; photons emitted roughly halfway across the ring, as seen from Earth, experience a large scattering optical depth along the line of sight. One way of improving the modeling of the N V $\lambda 1240$ light curve could therefore be to make the curvature of the bipolar nebula at its waist less extreme than in BL8, perhaps more like Figure 2 of Martin & Arnett (1995). In Martin & Arnett's model we estimate that the temperature and zero velocity optical depth are $(0.8 - 1.0) \times 10^5$ K and $15 - 55$, respectively, where the lower limits are for 500full1 and the upper for 500full2. The velocity shift is only $\sim 14.5$ km s$^{-1} \approx 1.4$ Doppler widths, so the value of $\tau_{lc}$ along the line of sight to the near side of the ring is $2 - 8$. This range is on the high side compared to what we used in the best fit cylinder models, but may be lower if recombination is important for the scattering gas, or if the column density along the line of sight is significantly less than in BL8. In BL8 the boundary between the N V and N IV zones occurs roughly halfway through the shocked RSG wind in this direction. In a situation with a lower column density, this boundary may occur outside the shocked RSG wind, which results in a lower optical depth in N V $\lambda 1240$ compared to Table 2, or BL8. Contrary to this, it is unlikely that recombination is important for the optical depth at large latitudes. This was demonstrated above for BL8, and the same result possibly applies to the models of Martin & Arnett, which have lower mass loss rates for the RSG wind. For recombination to become important, the density should be higher than a few $\times 10^2$ cm$^{-3}$ (Table 2). A density this high would imply that the scattering gas would contribute too much to the late line emission.

As demonstrated above, the IW model can qualitatively account for the resonance scattering of the N V $\lambda 1240$ line. Conversely, it is evident that the *the N V $\lambda 1240$ light curve can be used to distinguish between different hydrodynamic simulations*. It appears as if the model in Figure 2 of Martin & Arnett (1995) may explain the N V light curve better than BL8, but detailed calculations, like the one above for BL8, are needed to compare this and other models to BL8. We emphasize that the modeling of the N V $\lambda$1240 light curve, and the continuum observations of Crotts et al. (1995), are so far the best probes for the structure of the CSM in the region between the inner and outer rings. The very presence of the scattering gas, as well as the outer rings, seem difficult to explain in a natural way in the circumstellar disk model of McCray & Lin (1994).

## 4. Discussion

### 4.1. Abundances

In §3.3 we found that the relative abundances are insensitive to the density structure of the ring, as well as to whether we use 500full1 or 500full2 (Ensman & Burrows 1992) for the ionizing burst. Provided the general behavior of the evolution of the ionizing flux is not different from those models, this indicates that the abundances we derive are accurate to within the uncertainties of the atomic data and the observations. However, a different evolution of the burst could set up a different ionization and temperature structure within the ring (Lundqvist 1992), resulting in different relative line fluxes, and therefore different relative abundances from what we obtain in §3.3. We cannot exclude this possibility, but note that the ionization and temperature structures of the ring cannot be very different from that set up by the 500full1 and 500full2 bursts, because this would otherwise result in N V : N IV] : N III] : [N II] ratios different from the observed. We therefore believe that the C, N and O abundances we derive in §3.3 are accurate to $20 - 30\%$, and that the relative abundances of N/C and N/O are accurate to $\sim 40\%$, especially when the optical [N II] and [O III] lines are included, and not just those in the UV. A discussion of the abundances of other elements is given in Lundqvist et al. (1996).

The relative abundances of C, N and O we derive are in good agreement with what is found by Sonneborn et al. (1996), applying a nebular analysis. They obtained a slightly higher nitrogen abundance relative to carbon and oxygen by factors 1.2 and 1.5, respectively. Some of this discrepancy could be due to different atomic data. However, the main difference is probably a result of that Sonneborn et al. have to assume a temperature and ionization structure of the emitting gas, whereas this is calculated self-consistently in our models. In addition, we include light echo effects. Because of this, and the fact that the relative abundances we derive seem to be insensitive to the combination of parameters we have considered in our models, we believe our estimates are mainly limited by uncertainties of the atomic and observational data. Our estimates are therefore likely to have smaller systematic errors than those from the nebular analysis.

LF91 derive the same N/C ratio as we do, but obtain N/O $\sim 2$, which is higher than our estimate because of their older atomic data for the O III] $\lambda 1665$ line, and also because of a different ionization structure since in LF91 it was assumed that the gas was residing in a thin spherical shell.

In §3.2 we mentioned that the He/H ratio is uncertain mainly because of uncertain collision strengths of hydrogen. Collision strengths, and their temperature dependence, are likely to be uncertain also for He II. This may be the explanation to why the shape of the He II $\lambda 1640$ light curve is not modeled well. Another factor is that the He II line is weak, and therefore has a higher observational uncertainty. This would also explain why the weak O III] $\lambda 1665$ and N IV] $\lambda 1486$ lines are not modeled as well as, e.g., C III] $\lambda 1909$. To estimate the He/H ratio with better accuracy, information from optical helium lines should be included. This is done in Lundqvist et al. (1996).

### 4.2. Structure of the CSM

#### *4.2.1. The Inner Ring*

From the *HST* observations of the ring (e.g., Burrows et al. 1995) it is evident that our assumption of no azimuthal variation of the mass and/or density around the ring is a simplification. The ring appears clumpy, with structures down to the resolution of the images ($\sim 10^{17}$ cm; Burrows et al. 1995), and there is no reason to believe that there may not be even smaller structures. On a larger scale, the blue (near) side of the ring is in general brighter (Plait et al. 1995) and apparently less dense (Cumming et al. 1996), than the red. This could affect somewhat the modeled light curves. Assuming the azimuthal mass distribution of the gas emitting the UV lines to follow the brightness distribution of the *HST* image in [O III] on day 1278, Plait et al. (1995) showed that this does not change significantly the shape of the UV line light curves compared to a smooth ring. The difference occurs during the first $\sim 400$ days when the light echo paraboloid sweeps over the ring, and is negligible at subsequent epochs when recombination governs the shape of the light curves.

The *HST* images, and the line profile analysis by Cumming et al. (1996) trace the distribution of the gas observed at $t \geq 1144$ days. From our models we find that at this epoch gas with a density of $\lesssim 1.0 \times 10^4$ cm$^{-3}$ dominates the UV line emission. During the first $\sim 400$ days the corresponding density is $(2.0 - 3.5) \times 10^4$ cm$^{-3}$ (§3.3). The distribution of this gas may be different from the distribution of the less dense gas. We have therefore refrained from using an azimuthal distribution based on late ($t \geq 1144$ days) observations when modeling the UV line light curves, instead using a smooth distribution. Considering the results of Plait et al. (1995), this will not affect our conclusions.

Our models produce spikes in the light curves around day 410 when the light echo paraboloid has its last contact with the ring. These spikes are produced mainly by the high-density component. The near-absence of such spikes in the observed He II and N IV] light curves could therefore be due to that the ring is devoid of high-density material on the very far side of the ring. It may also be due to the low signal-to-noise which may swamp any spikes in the light curves of these lines. Even at low density, the N V line could show a spike, as seen in the low-density model in Figure 2, if the distribution of gas in the ring is smoother at low densities ($n \lesssim 10^4$ cm$^{-3}$).

The calculations in §3 show that for a smooth ring the opening angle, $\theta$ (see Fig. 6), is a few degrees, corresponding to a thickness perpendicular to the plane of the ring of $\approx 3.3 \times 10^{16}$ $(\theta/3°\!.0)$ cm, and that the radial thickness of the H II region is $(1-2) \times 10^{16}$ $(n/2.0 \times 10^4$ cm$^{-3})^{-1}$ cm. The radial thickness of the zone emitting the [O III] $\lambda\lambda 4959, 5007$ and the UV lines is even smaller, especially at late time (see Fig. 1). This can be compared with the observed projected thickness of the ring, which is $\gtrsim 10^{17}$ cm (e.g., Burrows et al. 1995; Plait et al. 1995). As already stated by Plait et al. (1995), the much smaller volume of emitting gas required by the photoionization calculations, compared to that inferred from the observations, argues for a low ($\ll 1$) filling factor of the emitting gas.

The filling factor is low either because the ring is just not fully resolved, or because it actually has an intrinsically low filling factor. One way to distinguish between these two cases is to study the azimuthal variation of the projected thickness around the ring. If the projected thickness is set mainly by the resolution, it should show very little variation with azimuth. Indeed, this is what is observed. If the ring were fully resolved, the small azimuthal variation would imply that the intrinsic height and radial extent of the ring must be about the same. If the radial extent dominates the projected thickness, and the ring is fully resolved, the thickness of the observed ellipse would be larger along its major axis, whereas if the thickness perpendicular to the plane of the ring dominates, the projected thickness would be thicker along the minor axis (see also Sonneborn et al. 1996). For the model in Figure 5 for 500full1, the intrinsic thickness of the ring is $\sim 4.4 \times 10^{16}$ cm, and the radial extent of the [O III] emitting gas is about a factor of two lower. The thickness perpendicular to the ring plane and the radial extent must therefore be of the same order, and from the small azimuthal variation of the projected thickness it appears impossible to judge whether the ring is fully resolved by the *HST* or not. In either case, the filling factor of the observed structure has to be $\sim 0.1$.

Plait et al. (1995) approximated the region around the waist of the nebula in the IW model by a torus with crescent-shaped cross section (see their Fig. 8), and found that this geometry gives a good fit to the radial profile of the [O III] emission, and may also provide an explanation for the low filling factor discussed above; the curvature of the nebula produces a skin of ionized gas, which, in projection, appears to occupy a large volume. The filling factor within the skin itself may be close to unity, which does not exclude that there may be some density variations around the ring (see above).

In §3.4 we found that the ionized part of the ring has a mass of $\sim 4.5 \times 10^{-2}$ M$_\odot$. We may interpret this as the swept up RSG wind in the IW model. Blondin & Lundqvist (1993) and Martin & Arnett (1995) argue that the last BSG phase lasted $\sim 2.0 \times 10^4$ yrs, so the ionized mass of the ring corresponds to a mass loss rate of the RSG in the equatorial plane of $\sim 2.2 \times 10^{-6}$ M$_\odot$ yr$^{-1}$. This is about an order of magnitude lower than in the models of Blondin & Lundqvist (1993) and Martin & Arnett (1995). In the IW scenario it is therefore only the inner part of the swept up RSG wind which is ionized by the burst, at least close to the equatorial plane. This agrees with our result in §3 that the ring is ionization bounded.

### 4.2.2. The Extended Nebula

A structure similar to what we infer for the gas scattering the N V $\lambda1240$ line was observed in the continuum by Crotts et al. (1995). Crotts et al. did not detect line emission from this gas, but later, weak lines were observed using the *HST* (Burrows et al. 1995). Because dust scattering, which is responsible for the continuum emission, scales as $\propto n$, and line emission scales roughly as $\propto n^2$, these observations are consistent with our result in §3.5 that the density of the gas below and above the equatorial plane must be considerably lower than in the ring. As stated in §3.5, the density should not be higher than $\sim 10^2$ cm$^{-3}$.

It is interesting to compare this limit to the density in the outer rings. A hint comes from the fact that the outer rings seem to fade roughly at the same rate as the inner ring (§1). From Table 2 it is obvious that the temperature and degree of ionization are much lower in the outer rings from the outset than in the inner ring. Because recombination and cooling proceed faster at lower temperatures and degrees of ionization (e.g., LF91), this implies that the density in the outer rings must be lower than in the component with the lowest density in the inner ring ($\sim 6 \times 10^3$ cm$^{-3}$). Using the results in §3.5, we estimate that the density of the gas in the outer rings observed by the *NTT* during the first $\sim 2000$ days should be a few $\times 10^3$ cm$^{-3}$. Detailed observations of the outer rings are needed to confirm this.

In §§3.5 and 4.2.1 we demonstrated that the IW model can account for an ionization bounded, circular mass concentration similar to the observed inner ring. We also argued that it provides an explanation to the gas necessary to scatter the N V $\lambda 1240$ line, and that it thus gives a good description of the density distribution up to a polar angle of $\sim 45°$, as viewed from the supernova. Martin & Arnett (1995) showed that the IW model is also able to qualitatively explain the outer rings, but Burrows et al. (1995) argued that the intensity contrast between the outer rings and the emission close to the rings is too large to be explained in terms of limb brightening, as in the IW model. Related to this is the fact that the high density we infer for the outer rings seems to be incompatible with the IW model; it appears as if the outer rings are material rings rather than a projection effect. Clearly, refinements to the IW model are needed to explain both the details of the N V $\lambda 1240$ light curve, and the density structure of the observed gas. An interesting extension to the standard IW model is the inclusion of photoionization by the progenitor of the CSM during the shaping of the nebula. Chevalier & Dwarkadas (1995) argue that this may produce the outer rings in a natural way. However, a detailed comparison with observations for this and other hydrodynamic models, like the CW model, cannot be assessed before detailed 2-D simulations have been made.

## 5. Conclusions

We have modeled the narrow emission lines from SN 1987A observed by the *IUE* and from the ground, up to day $\sim 2000$. The lines are emitted by gas associated with the inner ring observed by *NTT* and *HST*. To model the light curves of the lines we have used a time dependent photoionization code to follow the initial ionization of the ring during the supernova breakout, and the subsequent phase of recombination and cooling. From this we draw the following conclusions:

1. We are able to model the light curves of the lines in detail. The turn on around day 70, the time of maximum, and the shapes of the light curves are modeled well. The success of this, compared to LF91, stems from the ring geometry, in combination with refined models for the evolution of the ionizing radiation (Ensman & Burrows 1992). For a spherical shell, as in LF91, the population of low- and medium-ionization stages has to rely on recombination, causing a spread in the turn ons for the different lines, contrary to the observed turn ons of the UV lines, which occur nearly simultaneously. We find that the latter is a natural outcome of the ring geometry, since the ring is optically thick to the ionizing photons, creating broad C III, N III and O III zones from the outset. That the inner ring is ionization bounded implies that the observed ring is only the inner part of a more radially extended structure.

2. To account for the N V $\lambda$1240 light curve, resonance scattering in a medium external to the ring is needed. There is no need to include an X-ray tail as the result of the formation of a viscous shock shortly after the outburst to account for the N V $\lambda$1240 light curve, contrary to the suggestion by Luo (1991). The characteristic size of the scattering region, perpendicular to the plane of the ring, is of order $10^{18}$ cm. The structure we derive is similar to that inferred from the continuum observations of Crotts et al. (1995). We find that the resonance scattering of the N V line is a useful test of different hydrodynamical models for the formation of the circumstellar structure. In particular, the structure deduced from the N V line scattering is qualitatively consistent with the interacting winds model (e.g., Blondin & Lundqvist 1993; Martin & Arnett 1995). For a quantitative agreement more refined hydrodynamical models are necessary.

3. That the ring is ionization bounded agrees with the interacting winds model. From the results of Blondin & Lundqvist (1993) and Martin & Arnett (1995), we estimate that $\sim 10\%$ of the mass concentration in the equatorial plane in the interacting winds model is ionized by the burst. Our calculations show that the structure observed by the *HST* (e.g., Jakobsen et al. 1994; Burrows et al. 1995; Plait et al. 1995) requires that the ring has an apparent filling factor of $\sim 0.1$. This could be due to the limited resolution of *HST*, or that the ring has an intrinsically low filling factor. The latter could be a projection effect, if the geometry close to the equatorial plane is similar to that in the interacting winds model.

4. The mass of the ring observed up to $\sim 1900$ days is $\sim 4.5 \times 10^{-2}$ $M_\odot$. Though the ring is likely to contain gas with a range of densities, we find that we can fit the observed light curves satisfactorily with only three components with densities $6 \times 10^3$ cm$^{-3}$, $1.4 \times 10^4$ cm$^{-3}$ and $3.3 \times 10^4$ cm$^{-3}$. About half of the mass is confined to the component with the lowest density, while the other two components are about equal in mass. The component with the highest density dominates the light curves during the first $\sim 410$ days. There is then a successive shift to lower densities for the component dominating the light curves.

5. The most detailed models for the burst so far (Ensman & Burrows 1992) result in good fits to the light curves of the lines. From these calculations we find that the peak effective temperature was in the range $(5-8) \times 10^5$ K. The corresponding color temperature in these models is $(1.0-1.5) \times 10^6$ K. The spread in time of the burst due to different light travel times of photons emerging from the center compared to the limb of the supernova is found to be unimportant for the light curves.

6. We find N/C= $5.0 \pm 2.0$ and N/O= $1.1 \pm 0.4$ for the inner ring. The dominating

sources of uncertainty are the absolute fluxes derived from the observations, and the atomic data. The abundance of C+N+O is $0.30 \pm 0.05$ times solar, and He/H= $0.25 \pm 0.05$.

To establish reliable light curves for the optical lines, more optical data are needed than have been published during the first $\sim 3000$ days. Unfortunately, much of the observed optical data remains unpublished (e.g., Meurer et al. 1991; Wang et al. 1992; Caldwell et al. 1993). The optical and IR lines are mainly low-ionization lines, and therefore the only probes of the outer region of the ionized part of the inner ring. This region cannot be probed by the UV lines. From more comprehensive optical data, especially the flux of the Balmer lines, we may also establish a firmer He/H-ratio, as well as determine abundances of elements not emitting the strong UV lines observed by the *IUE*, e.g., Ne, S, Ca and Fe. Flux calibrated optical spectra also serve as a consistency check on the fluxes from the *IUE*. In particular, the comparison between [O III] $\lambda\lambda 4959, 5007$ and O III] $\lambda 1665$. Future observations of the inner ring are essential to determine the distribution of components with $n \lesssim 6 \times 10^3$ cm$^{-3}$. Continued observations of the outer rings are needed to determine their density distribution, which will constrain models for the formation of the circumstellar structure. For all observations, we emphasize that to get reliable light curves, continued monitoring of the same regions of the circumstellar gas is more important than observations of regions scattered over the nebula.

We thank John Blondin, Roger Chevalier, Robert Cumming, Ding Luo and Phil Plait for discussions, and George Sonneborn for helpful discussions and early access to data. Furthermore, we acknowledge support from the Swedish Natural Science Research Council and the Swedish Board of Space Research. C.F. also acknowledges support from the Göran Gustafsson Foundation for Research in Natural Sciences and Medicine.

## A. Atomic Data

Most of the atomic data we have used are the same as in LF91. However, many of the model atoms have been extended to include more levels, and we now also include silicon. For hydrogen we include levels up to $n = 7$, taking into account collisions between the sublevels of the five lowest $n$-states. For the $1s - 2s$ and $1s - 2p$ transitions we use the results of Scholz et al. (1990), and for collisions between $1s$ and $n = 3$ we adopt the results of Callaway & Unnikrishnan (1993). For the remaining transitions we use the results of Aggarwal et al. (1991a), except for collisions between $1s$ and $n = 4$, where we use mean values of the collision strengths of Aggarwal et al. (1991a) and Giovanardi, Natta, & Palla (1987). The former are larger by up to an order of magnitude. The collision strengths of Aggarwal et al. are generally too high due to the omission of continuum channels (Odgers, Scott, & Burke 1994), while the values of Giovanardi et al. are too low (Chang, Avrett, & Loeser 1991). The adopted values for the $1s$ to $n = 4$ transitions have profound effects on the strength of the Balmer lines, and hence the He/H ratio. Clearly, more accurate collision strengths are needed. For collisions among the $l$-states of all seven $n$-states, we use the calculations by Brocklehurst (1971). For He I we use the results by Almog & Netzer (1989), and for He II we consider the four lowest $n$-states with sublevels, using collision strengths by Aggarwal et al. (1991b) and the results by Brocklehurst (1971). In total, our hydrogen and helium model atoms contain 29 and 27 levels, respectively.

C I, N II, O III, Ne V and S III are treated as six-level atoms, and for C II, N I, N III, O I, O II, O IV, Ne III, Ne IV, Ne V, S I, S II, S IV and Si II we include five levels. Collision strengths and transition probabilities for these and other ions, except H I, He I and He II, were taken mainly from the compilations by Mendoza (1983), Gallagher & Pradhan (1985) and Aggarwal et al. (1986). We have updated these data by more recent calculations for C I (Johnson, Burke, & Kingston 1987), C II (Blum & Pradhan 1992), C III (Seaton 1987), N II (Stafford et al. 1994b), N III (Fang, Kwong, & Parkinson 1993; Stafford, Bell, & Hibbert 1994a), N IV (Keenan et al. 1986), O I (Berrington 1988), O II (McLaughlin & Bell 1993), O III (Aggarwal 1993), O IV (Blum & Pradhan 1992), Si II (Calamai, Smith, & Bergeson 1993), Si VI (Mohan & Le Dourneuf 1990) and S II (Cai & Pradhan 1993; Keenan et al. 1993). For the important coolant Fe II we calculate for $T \leq 3 \times 10^3$ K the populations of the 16 lowest levels using data from Nussbaumer & Storey (1980) and Berrington et al. (1988). For $T > 3 \times 10^3$ K we add the cooling due to transitions to, and among, higher levels using precalculated results for a 116 level atom. For Fe III and Fe IV we consider 30 and 22 levels, respectively, taking collision strengths from the compilations above. For Fe VII we assume nine levels, adopting the results by Keenan & Norrington (1987). In order to calculate cooling at low temperatures, we have included excitations to the fine structure levels of C II (Launay & Roueff 1977) and Si II (Roueff 1990) by collisions with H I.

We have also updated our recombination rates. For direct radiative recombination we include the low temperature results of Martin (1988) for H I, as well as the rates of Chung, Lin, & Lee (1991) for O I. For dielectronic recombination we substitute our rates for S I – S IV for $T > 4 \times 10^4$ K in LF91 by those calculated by Badnell (1991), however retaining our guessed form for the temperature dependence at lower temperatures (LF91). For dielectronic recombination to C II, N III and O I – O V we include the results of Badnell & Pindzola (1989a,b). The low temperature rates for Si I – Si III are taken from Nussbaumer & Storey (1986), and those at higher temperatures from Shull & van Steenberg (1982). For Si IV we use the rates by Romanik (1988). For higher ionization stages of Si we consider only the high temperature rates by Shull & van Steenberg (1982), and for Fe we use the compilation by Arnaud & Raymond (1992). Unfortunately, no low temperature rates for iron exist. Due to the diffuse emission from the ring and ambient stars, C II, Si II, S II and Fe II are assumed not to recombine, and therefore act as coolants also at low temperatures. To simulate this, we have artificially decreased the recombination coefficients for these ions.

In LF91 we used the photoionization cross sections by Reilman & Manson (1989). We have updated these for C, N, O and Fe by Opacity Project Data, made available using the TOPbase archive (Cunto & Mendoza 1993). For Fe II we also include the results by Le Dourneuf, Nahar, & Pradhan (1993). Charge transfer rates are taken from Arnaud & Rothenflug (1988) and references therein. We have updated these by H I reactions with N V (Zygelman et al. 1992) and Fe III (Neufeld & Dalgarno 1987).

### B. Radiative Transfer

To follow the ionization of the ring we use the same averaging procedure over optical depth for the mean intensity as Luo (1991). This gives good agreement with the more exact adaptive mesh method used by Lundqvist (1992), but is computationally faster. Typically, we use 50 radial mesh points with finer zoning in the inner parts of the ring. We treat the diffuse emission by an on-the-spot method similar to that in LF91, improving it by averaging escape probabilities over all relevant energies in the recombination continua, instead of using that at the threshold energies. This increases the continuum escape somewhat, and lowers the ionization due to diffuse emission. For the lines too, we use an escape probability formalism, taking escape probabilities from Rees, Netzer, & Ferland (1989).

Table 1. Characteristics of different burst models

| Model[a] | $T_{peak}$[b] | > 13.6 eV[c] | > 54.4 eV[c] | > 100 eV[c] |
|---|---|---|---|---|
| 10L | $2.0 \times 10^5$ | $5.8 \times 10^{56}$ | $4.3 \times 10^{55}$ | $4.8 \times 10^{54}$ |
| 11E0.8 | $5.5 \times 10^5$ | $6.9 \times 10^{56}$ | $1.4 \times 10^{56}$ | $6.9 \times 10^{55}$ |
| 500full1 | $5.5 \times 10^5$ | $1.0 \times 10^{57}$ | $1.5 \times 10^{56}$ | $8.2 \times 10^{55}$ |
| 500full2 | $7.3 \times 10^5$ | $2.3 \times 10^{57}$ | $3.2 \times 10^{56}$ | $1.6 \times 10^{56}$ |

[a]Models discussed are 10L (Woosley 1988), 11E0.8 (Shigeyama & Nomoto 1990), and 500full1 and 500full2 (Ensman & Burrows 1992).

[b]Peak effective temperature.

[c]Total number of ionizing photons with energy above 13.6 eV, 54.4 eV and 100 eV. Models have been joined to observations at 1.5 days which explains small differences for 500full1 and 500full2 compared to values given in Ensman & Burrows (1992). Note also that numbers for 10L and 11E0.8 differ from values given by LF91 and Ensman & Burrows (1992).

– 30 –Table 2.  Optical depth in N V $\lambda$1240 and temperature in the N V zone[a]

| Model[b] | $D$[c] | $\tau_{lc}$[d] | | | $T\ (10^5\ \text{K})$[e] | | |
|---|---|---|---|---|---|---|---|
| | ($10^{17}$ cm) | (1 day) | (10 days) | (100 days) | (1 day) | (10 days) | (100 days) |
| 500full1 | 6.3  | 165. | 122. | 16. | 1.59 | 1.58 | 1.22 |
| 500full1 | 9.0  | 76.  | 50.  | 1.8 | 1.31 | 1.28 | 0.76 |
| 500full1 | 16.0 | 10.  | 6.0  | …[f] | 0.77 | 0.73 | 0.48 |
| 500full1 | 25.0 | 1.0  | 0.58 | …[f] | 0.56 | 0.54 | 0.36 |
| 500full2 | 6.3  | 283. | 206. | 65. | 1.52 | 1.57 | 1.61 |
| 500full2 | 9.0  | 166. | 118. | 14. | 1.47 | 1.47 | 1.12 |
| 500full2 | 16.0 | 42.  | 26.  | 0.60 | 0.93 | 0.90 | 0.60 |
| 500full2 | 25.0 | 7.2  | 4.3  | …[f] | 0.69 | 0.65 | 0.44 |

[a]The ionized gas is ionization bounded and has density $2.0 \times 10^4\ \text{cm}^{-3}$.

[b]Models used for the outburst (Ensman & Burrows 1992).

[c]Distance of gas from the supernova.

[d]Optical depth through the N V zone at the line center of N V $\lambda$1240.

[e]Temperature at the peak of the radial distribution of the N V abundance.

[f]$\tau_{lc} < 0.10$. (Scattering of N V $\lambda$1240 is unimportant.)

– 32 –

- 35 -

– 37 –

Fig. 1.— Temperature and ionization as functions of the distance from the inner edge of the ring. Density is $n = 2.0 \times 10^4$ cm$^{-3}$, and the evolution of the burst is that of the 500full1 model by Ensman & Burrows (1992). The structure is shown at 2, 200 and 2000 days. Note the broad zones of N II, N III, O II and O III, and that these shift towards the inner side of the ring between days 2 and 200.

Fig. 2.— Line light curves for the model in Figure 1. The mass of the H II zone is $2.9 \times 10^{-2}$ M$_\odot$. Other parameters are given in the text. Observations of the UV lines are from Sonneborn et al. (1996), while the optical observations are from Wampler & Richichi (1989), Wampler, Richichi, & Baade (1989), Menzies (1991), Wang (1991), Khan & Duerbeck (1991) and Cumming (1994).

Fig. 3.— Same as Figure 2, but for $n = 2.0 \times 10^3$ cm$^{-3}$. The mass of the ionized gas is $5.6 \times 10^{-2}$ M$_\odot$. The low density was chosen to fit the N V $\lambda 1240$ light curve. Note, however, that with this density the light curves of the other lines cannot be modeled.

Fig. 4.— Same as Figure 2, but for 500full2. The mass of the H II zone is $3.1 \times 10^{-2}$ M$_\odot$.

Fig. 5.— Same as Figure 2, but for a mixed density model with $n = 6.0 \times 10^3$ cm$^{-3}$, $n = 1.7 \times 10^4$ cm$^{-3}$ and $n = 3.3 \times 10^4$ cm$^{-3}$. The total ionized mass is $4.5 \times 10^{-2}$ M$_\odot$. For N V $\lambda 1240$ resonance scattering in gas external to the ring is included (solid line). The dashed line gives the N V $\lambda 1240$ light curve without scattering.

Fig. 6.— Structure of the radial optical depth in N V $\lambda 1240$ for the BL8 model (Blondin & Lundqvist 1993; their model h) shortly after ionization. The 500full1 model by Ensman & Burrows (1992) was used for the supernova outburst. The N V $\lambda 1240$ emission mainly comes from the equatorial plane of the nebula, which subtends an angle $\theta$ at the supernova. A photon emitted towards Earth from point $A$ has a high probability to scatter at point $B$. Further scatterings are required to deflect it back towards Earth (path $BCDE$; path length $\Delta l$). The extra time spent in the nebula due to scattering is $\Delta l/c$. A photon emitted at $F$ towards Earth does not scatter because of the large velocity difference between the emitting gas and the gas at the near side of the nebula. The calculations discussed in the text allow for scattering in all three dimensions. The nebula is assumed to be tilted 45° to the line of sight.

Fig. 7.— N V $\lambda 1240$ light curves for models using the 500full1 and 500full2 bursts (Ensman & Burrows 1992). In the left hand column, the dashed light curve is the same as the N V $\lambda 1240$ light curve in Figure 5, whereas in the right hand column the dashed curve is for 500full2, the density components $n = 8.5 \times 10^3$ cm$^{-3}$, $n = 1.7 \times 10^4$ cm$^{-3}$ and $n = 3.3 \times 10^4$ cm$^{-3}$, and the total ionized mass $4.2 \times 10^{-2}$ M$_\odot$. Solid lines show the results when resonance scattering

in gas exterior to the ring is included. The upper panels are for a cylindrical structure of the scattering gas, whereas BL8 refers to the model in Figure 8 of Blondin & Lundqvist (1993). See text and Figure 6 for more details.